\documentclass[]{youtu} 

\usepackage{mathpazo}
\usepackage{graphicx}
\usepackage[numbers]{natbib}
\usepackage{float}
\usepackage{enumitem}
\usepackage{graphicx}
\usepackage[most]{tcolorbox}
\usepackage{subcaption}

\usepackage{tcolorbox}
\usepackage{enumitem}
\usepackage[dvipsnames]{xcolor}
\usepackage{fontawesome}
\usepackage{multirow}

\usepackage[utf8]{inputenc} 
\usepackage[T1]{fontenc}    
\usepackage{hyperref}       
\usepackage{url}            
\usepackage{booktabs}       
\usepackage{amsfonts}       
\usepackage{nicefrac}       
\usepackage{microtype}      
\usepackage{xcolor}         
\usepackage{amsmath}
\usepackage{enumitem}
\usepackage[most]{tcolorbox}
\usepackage{makecell}

\usepackage[utf8]{inputenc} 
\usepackage[T1]{fontenc}    
\usepackage{hyperref}       
\usepackage{url}            
\usepackage{booktabs}       
\usepackage{amsfonts}       
\usepackage{nicefrac}       
\usepackage{microtype}      
\usepackage{xcolor}         
\usepackage{amsmath}
\usepackage[most]{tcolorbox}

\makeatletter
\renewcommand\tableofcontents{%
  \@starttoc{toc}%
}

\let\origaddcontentsline\addcontentsline
\newcommand{\DisableTOC}{\let\addcontentsline\@gobblethree}
\newcommand{\EnableTOC}{\let\addcontentsline\origaddcontentsline}
\makeatother

\usepackage{booktabs}
\usepackage{makecell}
\usepackage{graphicx}
\usepackage{pifont}
\usepackage{adjustbox}
\usepackage{xcolor}
\usepackage{multirow}
\usepackage{caption}
\usepackage{calc} 
\usepackage{tabularx}
\newcommand{\vmark}{\textcolor{orange}{\textbullet}\ Video}
\newcommand{\imark}{\textcolor{teal}{\textbullet}\ Image}
\newcommand{\amark}{\textcolor{purple}{\textbullet}\ Audio}
\newcommand{\circleOpen}{\textcolor{blue}{\textbullet}\ Open}
\newcommand{\circleMC}{\textcolor{pink}{\textbullet}\ MC}
\usepackage[table]{xcolor}

\definecolor{headerblue}{RGB}{25, 60, 130}      
\definecolor{rowblueA}{RGB}{245, 248, 255}      
\definecolor{rowblueB}{RGB}{235, 242, 252}      
\definecolor{highlightblue}{RGB}{210, 225, 250} 
\definecolor{checkcolor}{RGB}{40, 120, 200}     
\definecolor{crosscolor}{RGB}{160, 170, 180}    

\newcommand{\cmark}{\textcolor{checkcolor}{\ding{51}}}
\newcommand{\xmark}{\textcolor{crosscolor}{\ding{55}}}

\definecolor{tableblue}{RGB}{205, 220, 240}
\definecolor{tablegreen}{RGB}{210, 240, 210} 
\definecolor{tablepink}{RGB}{245, 215, 225}

\newcommand{\bc}[1]{\cellcolor{blue!#1}}  
\newcommand{\gc}[1]{\cellcolor{green!#1}} 
\newcommand{\pc}[1]{\cellcolor{red!#1}}   

\usepackage{arydshln} 
\definecolor{myborder}{RGB}{73, 86, 102}
\definecolor{myRed}{RGB}{240, 48, 159}
\definecolor{mylightblue}{RGB}{235, 245, 255}

\tcbuselibrary{skins}

\title{Omni-DeepSearch: A Benchmark for Audio-Driven Omni-Modal Deep Search}
\author{
Tao Yu\textsuperscript{$1,2,3 * \S \spadesuit$}, 
Yiming Ding\textsuperscript{$1 * \dag$},
Shenghua Chai\textsuperscript{$1 \dag$}, 
Minghui Zhang\textsuperscript{$1 \dag$}, 
Zhongtian Luo\textsuperscript{$1 \dag$}, \\
Xinming Wang\textsuperscript{$1,2$}, 
Xinlong Chen\textsuperscript{$1,2$}, 
Zhaolu Kang\textsuperscript{$4$}, 
Junhao Gong\textsuperscript{$4$}, 
Yuxuan Zhou\textsuperscript{$5$}, \\
Haopeng Jin\textsuperscript{$1 \dag$}, 
Zhiqing Cui\textsuperscript{$1 \dag$}, 
Jiabing Yang\textsuperscript{$1,2$}, 
YiFan Zhang\textsuperscript{$1,2$}, , 
Hongzhu Yi\textsuperscript{$2 \ddagger$}, \\
Zheqi He\textsuperscript{$3 \ddagger$}, 
Xi Yang\textsuperscript{$3$}, 
Yan Huang\textsuperscript{$1,2 \ddagger$}, 
Liang Wang\textsuperscript{$1,2$}
}

\vspace{-5mm}

\affiliation{
\textsuperscript{$1$}CASIA \ 
\textsuperscript{$2$}UCAS \ 
\textsuperscript{$3$}BAAI
\textsuperscript{$4$}Peking University \
\textsuperscript{$5$}Tsinghua University
}

\date{May 9, 2026}
\sourcecode{https://github.com/yutao1024/Omni-DeepSearch}
\data{https://huggingface.co/datasets/Kirito-Lab/Omni-DeepSearch}

\begin{document}

\DisableTOC

\abstract{Current omni-modal benchmarks mainly evaluate models under settings where multiple modalities are provided simultaneously, while the ability to start from audio alone and actively search for cross-modal evidence remains underexplored. In this paper, we introduce \textbf{Omni-DeepSearch}, a benchmark for audio-driven omni-modal deep search. Given one or more audio clips and a related question, models must infer useful clues from audio, invoke text, image, and video search tools, and perform multi-hop reasoning to produce a short, objective, and verifiable answer. Omni-DeepSearch contains 640 samples across 15 fine-grained categories, covering four retrieval target modalities and four audio content types. A multi-stage filtering pipeline ensures audio dependence, retrieval necessity, visual modality necessity, and answer uniqueness. Experiments on recent closed-source and open-source omni-modal models show that this task remains highly challenging: the strongest evaluated model, Gemini-3-Pro, achieves only 43.44\% average accuracy. Further analyses illustrate key bottlenecks in audio entity inference, query formulation, tool-use reliability, multi-hop retrieval, and cross-modal verification. These results highlight audio-driven omni-modal deep search as an important and underexplored direction for future multimodal agents.}
\maketitle

\renewcommand{\thefootnote}{*}
\footnotetext{Equal contribution.}

\renewcommand{\thefootnote}{\dag}
\footnotetext{Work done during an internship at CASIA.}

\renewcommand{\thefootnote}{\S}
\footnotetext[0]{Work done during an internship at BAAI.}

\renewcommand{\thefootnote}{\textdaggerdbl}
\footnotetext{Corresponding author.}

\renewcommand{\thefootnote}{\ensuremath{\spadesuit}}
\footnotetext{Project leader.}
\renewcommand{\thefootnote}{\arabic{footnote}}

\vspace{-.1em}

\section{Introduction}

Humans frequently search for answers starting from sound: identifying a song from a melody, recognizing a speaker from a voice clip, or inferring a scene from ambient noise \citep{o2009world,dannenberg2007comparative,kabir2021survey,labourey2015sound}. Such processes often require more than audio recognition. Auditory cues must be converted into searchable queries, connected with external knowledge, and verified through text, image, or video evidence. However, current multimodal evaluation still lacks a systematic benchmark for this ability: starting from audio alone, actively searching in the open world, and reasoning across heterogeneous modalities.

\begin{table}[h]
\centering
\caption{Comparison of Omni-DeepSearch with existing benchmarks. Modality indicates the modalities involved in each benchmark. Multi-audio Input denotes whether multiple audio inputs are supported. Diverse Audio Categories specifies whether the input audio contains different types. Multi-Domain refers to the inclusion of data from a variety of real-world domains. Web-based Image or Video Search denotes whether online image or video search is included. Answer Type indicates the type of model responses, including open-ended and multiple-choice.}
\label{tab:benchmark_comparison}
\resizebox{\textwidth}{!}{
\rowcolors{2}{rowblueA}{rowblueB}
\begin{tabular}{llcccccccc}
\toprule

\hiderowcolors 
\rowcolor{headerblue}
{\color{white}\small\textbf{Benchmark}} & 
{\color{white}\small\textbf{Modality}} & 
{\color{white}\small\textbf{\makecell{Multi-audio\\Input}}} & 
{\color{white}\small\textbf{\makecell{Diverse Audio \\Categories}}} & 
{\color{white}\small\textbf{\makecell{Multi-hop\\Reasoning}}} & 
{\color{white}\small\textbf{\makecell{External\\Tools}}} & 
{\color{white}\small\textbf{\makecell{Multi-\\Domain}}} & 
{\color{white}\small\textbf{\makecell{Web-based\\Image Search}}} & 
{\color{white}\small\textbf{\makecell{Web-based\\Video Search}}} & 
{\color{white}\small\textbf{Answer Type}} \\
\showrowcolors 

\midrule

GAIA~\citep{mialon2023gaiabenchmarkgeneralai} & \imark & \xmark & \xmark & \cmark & \cmark & \cmark & \xmark & \xmark & \circleOpen \\
OmniBench~\citep{omnibench} & \imark / \amark & \xmark & \cmark & \xmark & \xmark & \cmark & \xmark & \xmark & \circleMC \\
AV-Odyssey~\citep{gong2024avodysseybenchmultimodalllms} & \imark / \amark & \cmark & \cmark & \xmark & \xmark & \cmark & \xmark & \xmark & \circleMC \\
WebWalkerQA~\citep{webwalker} & - & \xmark & \xmark & \cmark & \cmark & \xmark & \xmark & \xmark & \circleOpen \\
WorldSense~\citep{worldsense} & \vmark / \amark & \xmark & \cmark & \xmark & \xmark & \cmark & \xmark & \xmark & \circleMC \\
Daily-Omni~\citep{dailyomni} & \vmark / \amark & \xmark & \cmark & \xmark & \xmark & \xmark & \xmark & \xmark & \circleMC \\
BrowseComp-VL~\citep{webwatcher} & \imark & \xmark & \xmark & \cmark & \cmark & \cmark & \cmark & \xmark & \circleOpen \\
OmniVideoBench~\citep{li2026omnivideobenchaudiovisualunderstandingevaluation} & \vmark / \amark & \xmark & \cmark & \cmark & \xmark & \cmark & \xmark & \xmark & \circleMC \\
UNO-Bench~\citep{unobench} & \vmark / \imark / \amark & \xmark & \cmark & \cmark & \xmark & \cmark & \xmark & \xmark & \circleMC / \circleOpen \\
VideoBrowserComp~\citep{videobrowser} & \vmark & \xmark & \xmark & \cmark & \cmark & \cmark & \xmark & \cmark & \circleOpen \\
VideoDR~\citep{liu2026watchingreasoningsearchingvideo} & \vmark & \xmark & \xmark & \cmark & \cmark & \cmark & \xmark & \xmark & \circleOpen \\
EmoOmniEval~\citep{tian2026emoomnibridgingemotionalunderstanding} & \vmark / \amark & \xmark & \xmark & \cmark & \xmark & \cmark & \xmark & \xmark & \circleOpen \\
OmniGAIA~\citep{omnigaia} & \vmark / \imark / \amark & \xmark & \xmark & \cmark & \cmark & \cmark & \cmark & \xmark & \circleOpen \\
MMOU~\citep{goel2026mmoumassivemultitaskomni} & \vmark / \amark & \xmark & \xmark & \xmark & \xmark & \cmark & \xmark & \xmark & \circleMC / \circleOpen \\
SocialOmni~\citep{xie2026socialomnibenchmarkingaudiovisualsocial} & \vmark / \amark & \xmark & \xmark & \cmark & \xmark & \cmark & \xmark & \xmark & \circleMC / \circleOpen \\
HumanOmni-Speaker~\citep{bai2026humanomnispeakeridentifyingsaid} & \vmark / \amark & \xmark & \xmark & \xmark & \xmark & \xmark & \xmark & \xmark & \circleMC / \circleOpen \\
OmniACBench~\citep{kim2026omniacbenchbenchmarkevaluatingcontextgrounded} & \imark / \amark & \xmark & \cmark & \xmark & \xmark & \xmark & \xmark & \xmark & \circleOpen \\
OMD-Bench~\citep{nazi2026omnimodaldissonancebenchmarksystematically} & \vmark / \amark & \xmark & \cmark & \cmark & \xmark & \cmark & \xmark & \xmark & \circleMC \\
Video-to-Script~\citep{pu2026omniscriptaudiovisualscriptgeneration} & \vmark / \amark & \xmark & \cmark & \cmark & \xmark & \cmark & \xmark & \xmark & \circleOpen \\
AVID~\citep{zhang2024avidanylengthvideoinpainting} & \vmark / \amark & \xmark & \cmark & \xmark & \xmark & \cmark & \xmark & \xmark & \circleMC / \circleOpen \\

\midrule
\rowcolor{highlightblue}
\textbf{Ours} & \vmark / \imark / \amark & \cmark & \cmark & \cmark & \cmark & \cmark & \cmark & \cmark & \circleOpen \\

\bottomrule
\end{tabular}
}
\end{table}


As Deep Search has become an important paradigm for evaluating open-domain reasoning, existing benchmarks have expanded from text-based web search to image-text retrieval and video browsing, but they typically assume the initial clue is textual or visual~\citep{webwalker,webwatcher,videobrowser}. Audio remains largely underexplored as the origin of deep search, despite its ubiquity in real-world scenarios. This is challenging because models must understand ambiguous auditory signals and transform them into effective queries for cross-modal evidence gathering.

One may evaluate audio under joint multimodal inputs \citep{omnigaia,chen2026diademadvancingdialoguedescriptions}, but this does not provide a clean measure of audio-driven search ability. Prior studies show that models may over-rely on visual information when audio and visual signals are presented together~\citep{zhao2025multifacetedevaluationaudiovisualcapability,selvakumar2026audiovisuallargelanguagemodels}. Therefore, we use audio as the only initial modality, forcing models to infer useful clues from sound before invoking external text, image, and video search tools.


We introduce \textbf{Omni-DeepSearch}, a benchmark for audio-driven omni-modal deep search. In this benchmark, a model is given one or more audio clips together with a related question, and is required to search in open environments, gather external evidence, and generate a short, objective, and verifiable answer. Omni-DeepSearch contains 640 samples covering 15 fine-grained categories. These categories are defined along two axes: the target modality to be retrieved and the type of audio content provided as input. The retrieval targets include text search with a single audio input, text search with multiple audio inputs, image-text search with a single audio input, and video search with a single audio input. The audio inputs cover speech, ambient sound, music, and animal sounds. To ensure benchmark quality, we design a multi-stage filtering pipeline that verifies audio dependence, retrieval necessity, visual-modality necessity, answer uniqueness, and answer verifiability. Experiments on omni-modal models show that Omni-DeepSearch is highly challenging. The strongest evaluated model, Gemini-3-Pro, achieves only 43.44\% average accuracy, while open-source models lag substantially behind. Further analyses reveal several key bottlenecks, including audio entity inference, query formulation, tool-use reliability, multi-hop retrieval, and cross-modal verification. These findings suggest that audio-driven deep search requires more than isolated audio recognition: models must coordinate auditory perception with external search and multimodal reasoning.

Our main contributions are summarized as follows:

We identify and formalize audio-driven omni-modal deep search, where audio is the sole initial modality and models must actively search and reason across text, image, and video evidence.

We construct Omni-DeepSearch with 640 samples across 15 fine-grained categories, covering four retrieval target modalities and four audio content types. A multi-stage construction and filtering pipeline further ensures audio dependence, retrieval necessity, visual modality necessity, answer uniqueness, and reliable evaluation.

We conduct extensive experiments and analyses on recent omni-modal models, revealing key limitations in audio entity inference, query formulation, tool use, multi-hop retrieval, and cross-modal verification.





\section{Related Works}
\subsection{Omni-modal Evaluation Benchmarks}

Omni-modal benchmarks evaluate models' ability to integrate visual, auditory, and textual information. Existing work mainly focuses on joint perception and reasoning: OmniBench~\citep{omnibench} and UNO-Bench~\citep{unobench} test image-audio-text understanding, WorldSense~\citep{worldsense} and Daily-Omni~\citep{dailyomni} examine audio-visual temporal alignment in videos, and OmniGAIA~\citep{omnigaia} extends evaluation to tool-augmented multi-modal reasoning. However, these benchmarks typically provide all relevant modalities simultaneously, so the challenge lies in aligning co-present signals rather than discovering cross-modal evidence from a single modality. In contrast, Omni-DeepSearch uses audio as the only initial modality, requiring models to infer clues from sound and actively retrieve text, image, and video evidence for multi-hop reasoning. For specific differences, see Table \ref{tab:benchmark_comparison}.


\subsection{Deep Search Benchmarks}


Deep Search benchmarks evaluate models' ability to solve complex open-domain problems through multi-step retrieval, tool use, and iterative reasoning. Existing work has expanded from text-based web search to multimodal settings: WebWalker focuses on web navigation and textual information gathering~\citep{webwalker}, WebWatcher incorporates visual-textual evidence for vision-language deep search~\citep{webwatcher}, and Video-Browser extends deep search to video localization and fine-grained visual verification~\citep{videobrowser}. However, audio remains underexplored as the initial information source. Prior studies show that models may over-rely on visual inputs when audio and video are presented together~\citep{zhao2025multifacetedevaluationaudiovisualcapability,selvakumar2026audiovisuallargelanguagemodels}. In contrast, Omni-DeepSearch starts from audio alone and requires models to invoke text, image, and video search tools for cross-modal reasoning.

\section{Omni-DeepSearch Bench}
\begin{figure}[t]
    \centering
    \includegraphics[width=0.95\linewidth]{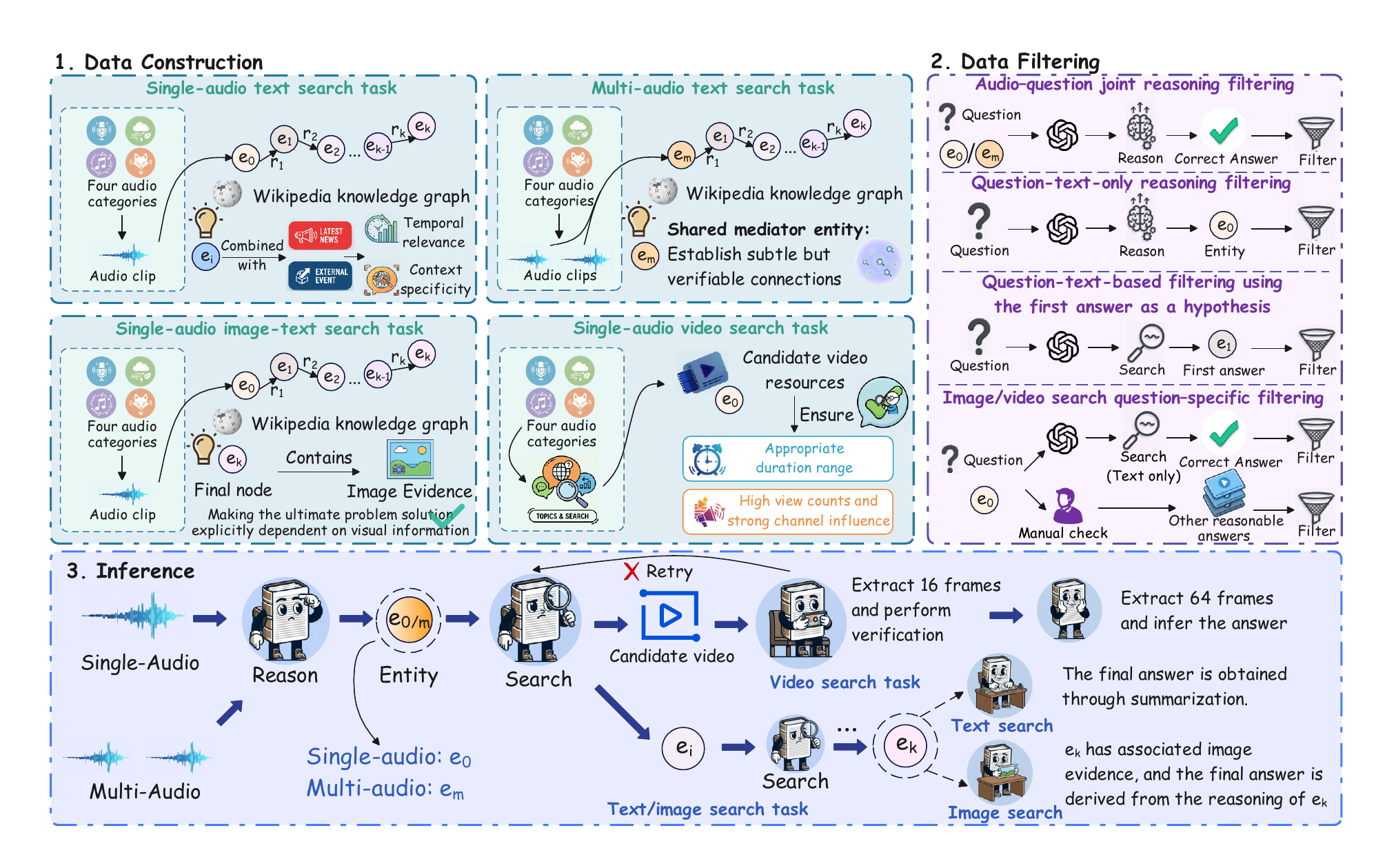}
    \caption{Overview of Omni-DeepSearch. In data construction, tasks are built across four audio categories and four retrieval settings. Text and image-text tasks are constructed over Wikipedia knowledge-graph paths, while video tasks are collected from filtered candidate videos. In data filtering, multi-stage LLM-based checks ensure audio dependence, retrieval necessity, visual modality necessity, and answer uniqueness. During inference, models start from audio alone, infer the audio-related entity, and iteratively invoke text, image, or video search tools to gather cross-modal evidence and produce the final answer.}
    \label{fig:1}
    \vspace{-10pt}
\end{figure}
\vspace{-10pt}







\subsection{Task Definition and Design Principles}

We introduce the \textbf{Omni-DeepSearch} task for audio-driven information retrieval. Given one or more audio clips and a related deep-search question, the model must use multimodal retrieval tools, including text search, image search, and video search, to gather evidence from open sources and produce a short, objective, and verifiable answer.

This task is designed around four principles. First, \textbf{mandatory audio dependence}: every question is anchored in the input audio, so it cannot be answered solely from the question text or prior knowledge. The model must infer key audio-related cues, such as speaker identity, acoustic events, music, or sound sources, before retrieval can begin. Second, \textbf{hard to find, easy to verify}: questions require multi-step search and reasoning, while answers are restricted to automatically comparable strings such as entity names, quantities, or colors. Third, \textbf{omni-modal retrieval}: although audio is the only initial modality, solving the task may require text, image, or video search, depending on the question. Fourth, \textbf{answer uniqueness}: each question has one definitive ground-truth answer based on objective and verifiable evidence.

These principles make Omni-DeepSearch challenging in ways that differ from conventional multimodal benchmarks. Audio cues are often ambiguous and not directly searchable, requiring models to convert uncertain auditory perception into effective queries. The resulting evidence chains can be long and fragile, especially in multi-audio tasks where models must infer a shared mediator entity from several clips. For image-text and video tasks, models must further perform fine-grained visual verification or temporal reasoning. Thus, Omni-DeepSearch evaluates not only audio understanding, but also query formulation, tool use, multi-hop retrieval, and cross-modal verification.

\subsection{Task Taxonomy}

To systematically evaluate cross-modal understanding, retrieval, and reasoning in the Omni Deep Search setting, we organize tasks along two orthogonal dimensions: \emph{retrieval target modality} and \emph{audio content type}. The former characterizes which external information sources the model must invoke to complete the deep search, while the latter characterizes the semantic properties and perceptual features of the input audio. Their combination yields a task space with well-defined structure and broad coverage.

\subsubsection{Retrieval Target Modality}

We divide Omni-DeepSearch tasks into four categories according to the retrieval modality required for solving the question.

\textbf{1. Single-audio text search.}
Given one audio clip and a related question, the model must infer key audio cues and answer the question through text search and multi-hop reasoning.

\textbf{2. Multi-audio text search.}
Given multiple audio clips and a unified question, the model must integrate complementary clues across clips, infer their shared connection, and complete the answer through text search.

\textbf{3. Single-audio image-text search.}
Given one audio clip and a question, the model must first identify the relevant entity or context from audio, then retrieve and verify image evidence together with textual information to derive the answer.

\textbf{4. Single-audio video search.}
Given one audio clip and a question, the model must retrieve the corresponding video and reason over its temporal visual content, requiring audio-to-video alignment and fine-grained video understanding.

\subsubsection{Audio Content Type}
\label{sec:audio_type}

Beyond retrieval modality, we classify input audio into four content types following~\citep{park2025natural,fonseca2021fsd50k}, as different audio signals pose different perception and retrieval challenges.

\textbf{1. Speech.}
Speech clips include speeches, interviews, dialogues, and narration, where key clues may lie in speaker identity, vocal characteristics, linguistic content, or contextual background.

\textbf{2. Ambient sound.}
Ambient clips contain natural or scene-level sounds, such as traffic, machinery, wind, rain, or urban soundscapes, requiring models to infer scene and sound-source information without explicit language.

\textbf{3. Music.}
Music clips contain melodies, rhythms, instruments, or vocal performances, requiring models to connect acoustic patterns with musical works, performers, styles, or cultural knowledge.

\textbf{4. Animal sound.}
Animal sound clips contain calls, roars, or other biological vocalizations, requiring models to identify species or sound classes and reason about related ecological or behavioral context.

\subsubsection{Combined Task Space}

We organize the 15 task categories as follows: for each of the three single-audio retrieval modalities, we cross with four audio content types (12 categories); multi-audio text search is further divided by the number of audio clips (3 categories).


\subsection{Dataset Construction}

\subsubsection{Single-Audio Text Search Tasks}

For \textbf{single-audio text search tasks}, we first collect audio clips from YouTube spanning the four categories introduced in Section~\ref{sec:audio_type}: speech, ambient sound, music, and animal sound. Questions are constructed over a \textbf{knowledge graph built from Wikipedia}, $\mathcal{G} = (\mathcal{E}, \mathcal{R})$, where $\mathcal{E}$ denotes the entity set and $\mathcal{R}$ the relation set. Each question is generated by constructing a path starting from an entity $e_0 \in \mathcal{E}$ directly associated with the audio clip, ensuring that the question is grounded in the audio content. The prompt is provided in Appendix \ref{single}. The construction proceeds as follows:

\textbf{Path construction.} Starting from the audio-associated entity $e_0$, we traverse graph relations to form a path of length $k \geq 5$: $p = (e_0 \xrightarrow{r_1} e_1 \xrightarrow{r_2} e_2 \cdots \xrightarrow{r_k} e_k), \quad e_i \in \mathcal{E},\ r_i \in \mathcal{R}.$ This ensures that answering the question requires multi-hop search and reasoning from the audio-grounded information, rather than single-step fact lookup.

\textbf{Retrieval dependency reinforcement.} To prevent models from leveraging parametric knowledge to directly infer inter-entity relations along the path and thereby bypass retrieval, we randomly select a node $e_i\ (i>0)$ in each sample and bind it to a recent news event during question generation. This introduces temporal specificity and contextual freshness, increasing both the realism and difficulty of the task.

Together, these two steps guarantee that every question is \textbf{audio-dependent}, demands multi-step reasoning, and yields a temporally grounded, verifiable answer.

\subsubsection{Multi-Audio Text Search Tasks}

For the \textbf{multi-audio text search task}, we first select a group of audio entities $\{e^a_1,\dots,e^a_n\}$ ($n \leq 4$) from different domains, corresponding to the four audio categories introduced in Section~\ref{sec:audio_type}: speech, music, animal sound, and ambient sound, with each entity possessing distinctive acoustic characteristics. We then identify a shared mediator entity $e_m$ in the Wikipedia-based knowledge graph $\mathcal{G}=(\mathcal{E},\mathcal{R})$ that is connected to all audio entities via graph relations, establishing a verifiable link among them. Audio clips for each entity are collected from YouTube. The prompt is provided in Appendix \ref{multi}. The construction proceeds as follows:

\textbf{Path construction.} The model is required to identify the respective audio entities from the multiple clips and further infer their shared mediator entity $e_m$. From $e_m$, we construct a multi-hop path of length $k \geq 5$ along the knowledge graph: $p=(e_m \xrightarrow{r_1} e_1 \xrightarrow{r_2} e_2 \cdots \xrightarrow{r_k} e_k),
\quad e_i \in \mathcal{E},\ r_i \in \mathcal{R}.$
This requires the model to first integrate information across multiple audio clips and then perform multi-hop retrieval and reasoning grounded in the shared mediator entity, increasing the compositional and reasoning difficulty of the task.

\textbf{Retrieval dependency reinforcement.} To prevent models from leveraging parametric knowledge to directly infer inter-entity relations along the path and thereby bypass retrieval, we randomly select a non-starting node $e_i$ in each sample and bind it to a recent news event during question generation. This introduces temporal specificity and contextual freshness, increasing both the realism and difficulty of the task.

Together, these two steps guarantee that every question is \textbf{audio-dependent}, demands cross-clip integration and multi-step reasoning, and yields a temporally grounded, verifiable answer.



\subsubsection{Single-Audio Image-Text Search Tasks}

For the \textbf{single-audio image-text search task}, we first collect audio clips from YouTube spanning the four categories introduced in Section~\ref{sec:audio_type}: speech, ambient sound, music, and animal sound. Questions are constructed over the same Wikipedia-based knowledge graph $\mathcal{G}=(\mathcal{E},\mathcal{R})$, starting from an entity $e_0 \in \mathcal{E}$ directly associated with the audio content. The prompt is provided in Appendix \ref{image}. The construction proceeds as follows:

\textbf{Path construction.} Starting from the audio-associated entity $e_0$, we traverse graph relations to form a path of length $k \geq 3$: $p=(e_0 \xrightarrow{r_1} e_1 \xrightarrow{r_2} e_2 \cdots \xrightarrow{r_k} e_k),
\quad e_i \in \mathcal{E},\ r_i \in \mathcal{R}.$
Unlike the single-audio text search task, this task requires that the final node $e_k$ be verifiable through external image evidence, making the ultimate answer explicitly dependent on visual information in addition to text retrieval and multi-hop reasoning.

\textbf{Visual verification.} The answer to each question cannot be determined by text alone; the model must locate a relevant image associated with $e_k$ and perform fine-grained visual inspection to produce the final answer. This ensures that the task evaluates not only audio comprehension and multi-hop retrieval but also cross-modal mapping from audio to visual evidence.

Together, these two steps guarantee that every question is \textbf{audio-dependent}, demands multi-step reasoning, and requires explicit visual verification.

\subsubsection{Single-Audio Video Search Tasks}

For the \textbf{single-audio video search task}, we first construct relevant themes and search queries around the four audio categories introduced in Section~\ref{sec:audio_type}: speech, ambient sound, music, and animal sound, and use them to retrieve candidate video resources. The prompt is provided in Appendix \ref{video}. The construction proceeds as follows:

\textbf{Video filtering.} To ensure data quality and source traceability, we impose duration and source reliability constraints on the candidate videos, retaining only those that fall within a specified duration range and exhibit high view counts and subscriber numbers.

\textbf{Question generation.} From the filtered videos, we extract audio clips that satisfy duration requirements, are semantically coherent, and are representative of the source video content. Each question requires the model to first retrieve the corresponding video from open-domain video resources given only the audio, and then perform fine-grained reasoning over the temporal visual content of the video.

Together, these two steps guarantee that every question is \textbf{audio-dependent} and demands both cross-modal video retrieval and temporal visual reasoning.

\subsection{Data Filtering}

To ensure data quality, we first use Gemini-3-Pro to review each extracted audio clip, retaining only clips with a clear dominant sound source, an unambiguous category, and consistency with the four audio types in Section~\ref{sec:audio_type}. Clips with strong noise, mixed sound sources, or category mismatches are removed.

For generated question samples, inspired by the MLLM-based filtering strategy in MM-DeepResearch~\citep{mm-deepresearch}, we use GPT-5 to perform multi-stage filtering tailored to the audio-driven setting. We denote the entity directly associated with the audio as the \emph{audio subject} ($e_0$ for single-audio tasks and $e_m$ for multi-audio tasks). The filtering includes four stages: (1) \textbf{joint audio-question reasoning}, which removes samples answerable from the audio subject and question without retrieval; (2) \textbf{question-only reasoning}, which removes samples whose audio subject can be inferred from the question alone; (3) \textbf{first-hop entity leakage filtering}, which removes samples where the first-hop entity $e_1$ can be retrieved using only the question text; and (4) \textbf{visual modality necessity filtering}, which removes image-text and video samples that can be answered with text search alone or admit multiple plausible visual answers.

This process ensures that final samples are audio-dependent, retrieval-demanding, modality-appropriate, and uniquely verifiable. The filtering prompts are provided in Appendix~\ref{filter}.



\subsection{Data Statistics}

The final benchmark dataset comprises 640 samples, spanning four retrieval task types and four audio content categories. Following the task space defined above, the dataset is further divided into 15 fine-grained task categories. The sample distribution across categories is shown in Figure ~\ref{fig:pie}.

\section{Experiments}

\subsection{Baseline}









The inference baseline follows the tool-augmented reasoning pipeline of~\citep{mm-deepresearch}. Given audio clips and a corresponding question, the model iteratively invokes external tools, including text search, image search, and video search, over multiple reasoning rounds to progressively gather cross-modal evidence and produce a final textual answer. The overall pipeline is illustrated in Figure~\ref{fig:1}. The prompt is provided in Appendix \ref{2}. The inference proceeds as follows:

\textbf{Audio comprehension and entity grounding.} The model first comprehends the input audio to identify the retrieval starting point. For single-audio tasks, it recognizes the entity $e_0$ directly associated with the audio content; for multi-audio tasks, it integrates information across multiple clips to infer the shared mediator entity $e_m$. The model then uses $e_0$ or $e_m$ as the initial cue to launch subsequent retrieval and reasoning.

\textbf{Multi-hop retrieval and answer derivation.} The subsequent retrieval strategy varies by task type:

\textit{Text search and image--text search tasks.} The model performs multi-hop search over open-domain retrieval results, progressively obtaining intermediate entities and evidence nodes $e_i, \dots, e_k$. For text search tasks, the answer is derived by synthesizing evidence gathered across multiple rounds of text retrieval. For image--text search tasks, the model further retrieves image evidence associated with the final node $e_k$ and reasons over the visual content to produce the answer.

\textit{Video search tasks.} The model retrieves candidate videos based on the identified entity, audio cues, and the question, and extracts a small number of keyframes (e.g., 16 frames) for rapid verification. If the candidate video does not match the audio or the question, the model re-queries or selects a new candidate; if verification succeeds, it extracts a denser frame sequence (e.g., 64 frames) to perform fine-grained reasoning over the temporal visual content and generate the final answer.

This unified pipeline enables evaluation of audio comprehension, multi-hop search, cross-modal evidence integration, and temporal visual reasoning within a single framework.

\subsection{Evaluation Metrics and Settings}

We use accuracy as the primary metric. Since answers in Omni-DeepSearch are short, objective, and uniquely verifiable, each prediction is judged against the ground truth using an LLM-based protocol. Three strong LLM judges, GPT-5.4~\citep{openai2026gpt54}, Gemini-3-pro~\citep{gemini3pro2025}, and Claude-Sonnet-4.6~\citep{anthropic2026sonnet46}, independently assess semantic equivalence, with the final label determined by majority vote.

We evaluate both closed-source and open-source models, including Gemini-3-Pro~\citep{gemini3pro2025}, Gemini-3-Flash~\citep{gemini3pro2025}, Gemini-2.5-Pro~\citep{comanici2025gemini}, Gemini-2.5-Flash-Lite~\citep{comanici2025gemini}, Qwen3.5-Omni-Plus/Flash~\citep{team2026qwen3}, Mimo-V2-Omni~\citep{xiao2026mimo}, Mimo-V2.5~\citep{mimov25}, Qwen3-Omni-30B-A3B~\citep{xu2025qwen3}, and Qwen2.5-Omni~\citep{xu2025qwen2}. We report the overall accuracy on 640 data instances, as well as accuracy by retrieval target modality and audio content type. Implementation details are provided in Appendix~\ref{canshu}.

\subsection{Main Results}





Table~\ref{tab:main_results} presents the main results on Omni-DeepSearch.

\textbf{Frontier models show clear advantages, but overall performance remains far from saturated.}
Gemini-3-Pro achieves the best performance among all evaluated models, with an average accuracy of 43.44\%, substantially outperforming both closed-source and open-source models. This indicates that frontier omni-modal models already possess a certain degree of audio-driven deep search capability. However, the overall accuracy remains far from saturated, suggesting that Omni-DeepSearch is still challenging even for the strongest evaluated model.

\textbf{Cross-modal retrieval significantly increases task difficulty.}
Models generally perform better on single-audio text search than on the other three task types. For example, Gemini-3-Pro obtains 57.50\% accuracy on single-audio text search, but drops to 40.63\%, 38.75\%, and 36.88\% on multi-audio text search, image--text search, and video search, respectively. This trend shows that the task becomes more difficult when models need to integrate multiple audio clips, verify visual evidence, or reason over temporal video content. In particular, video search is consistently challenging, as it requires both retrieving the correct video from audio cues and locating relevant visual evidence within the video.

\begin{table*}[t]
\centering
\caption{Experimental results of closed-source and open-source models on Omni-DeepSearch. Audio Content Type columns are from single-audio tasks (480 samples); Avg is over all 640.}
\label{tab:main_results}
\setlength{\tabcolsep}{4pt}
\renewcommand{\arraystretch}{1.2}
\resizebox{0.85\textwidth}{!}{%
\begin{tabular}{l 
                *{4}{>{\centering\arraybackslash}p{1.3cm}} 
                *{4}{>{\centering\arraybackslash}p{1.3cm}} 
                >{\centering\arraybackslash}p{1.3cm}}
\toprule
\multirow{2}{*}[-0.5ex]{\textbf{Model}} 
& \multicolumn{4}{c}{\textbf{Retrieval Target Modality}} 
& \multicolumn{4}{c}{\textbf{Audio Content Type}} 
& \multirow{2}{*}{\textbf{Avg}} \\
\cmidrule(lr){2-5} \cmidrule(lr){6-9} 
 & SINGLE & MULTI & IMAGE & VIDEO & SPEECH & MUSIC & BIO & ENV &  \\
\midrule
\multicolumn{10}{l}{\textbf{Closed Source}} \\
\hdashline 
\addlinespace[2pt]
Gemini-3-Pro                  & \bc{92} 57.50 & \bc{65} 40.63 & \bc{62} 38.75 & \bc{59} 36.88 & \gc{66} 55.00 & \gc{56} 46.67 & \gc{47} 39.17 & \gc{44} 36.67 & \pc{70} \textbf{43.44} \\
Gemini-3-Flash                & \bc{43} 26.88 & \bc{35} 21.88 & \bc{34} 21.25 & \bc{19} 11.88 & \gc{24} 20.00 & \gc{22} 18.33 & \gc{20} 16.67 & \gc{30} 25.00 & \pc{33} 20.47 \\
Gemini-2.5-Pro                & \bc{33} 20.62 & \bc{22} 13.75 & \bc{24} 15.00 & \bc{33} 20.62 & \gc{30} 25.00 & \gc{19} 15.83 & \gc{26} 21.67 & \gc{15} 12.50 & \pc{28} 17.50 \\
Gemini-2.5-Flash-Lite         & \bc{2} 1.25   & \bc{7} 4.38   & \bc{0} 0.00   & \bc{5} 3.13   & \gc{5} 4.17   & \gc{0} 0.00   & \gc{1} 0.83   & \gc{1} 0.83   & \pc{4} 2.19 \\
\addlinespace[1pt]
Qwen3.5-Omni-Plus             & \bc{32} 20.00 & \bc{15} 9.38  & \bc{25} 15.62 & \bc{17} 10.62 & \gc{17} 14.17 & \gc{19} 15.83 & \gc{18} 15.00 & \gc{20} 16.67 & \pc{22} 13.91 \\
Qwen3.5-Omni-Flash            & \bc{10} 6.25  & \bc{4} 2.50   & \bc{11} 6.88  & \bc{5} 3.13   & \gc{5} 4.17   & \gc{5} 4.17   & \gc{7} 5.83   & \gc{8} 6.67   & \pc{8} 4.69 \\
\addlinespace[1pt]
Mimo-V2-Omni                  & \bc{23} 14.38 & \bc{6} 3.75   & \bc{19} 11.88 & \bc{14} 8.75  & \gc{13} 10.83 & \gc{10} 8.33  & \gc{15} 12.50 & \gc{18} 15.00 & \pc{16} 9.69 \\
\midrule
\multicolumn{10}{l}{\textbf{Open Source}} \\
\hdashline 
\addlinespace[2pt]
Mimo-V2.5                     & \bc{24} 15.00 & \bc{15} 9.38  & \bc{23} 14.38 & \bc{13} 8.13  & \gc{19} 15.83 & \gc{8} 6.67   & \gc{19} 15.83 & \gc{14} 11.67 & \pc{19} 11.72 \\
\addlinespace[1pt]
Qwen3-Omni-30B-A3B (Thinking) & \bc{15} 9.38  & \bc{10} 6.25  & \bc{17} 10.62 & \bc{0} 0.00   & \gc{7} 5.83   & \gc{9} 7.50   & \gc{7} 5.83   & \gc{9} 7.50   & \pc{11} 6.56 \\
Qwen3-Omni-30B-A3B (Instruct) & \bc{11} 6.88  & \bc{4} 2.50   & \bc{3} 1.88   & \bc{4} 2.50   & \gc{2} 1.67   & \gc{6} 5.00   & \gc{6} 5.00   & \gc{4} 3.33   & \pc{6} 3.44 \\
Qwen2.5-Omni-7B               & \bc{3} 1.88   & \bc{0} 0.00   & \bc{1} 0.62   & \bc{3} 1.88   & \gc{2} 1.67   & \gc{0} 0.00   & \gc{3} 2.50   & \gc{2} 1.67   & \pc{2} 1.09 \\
Qwen2.5-Omni-3B               & \bc{2} 1.25   & \bc{1} 0.62   & \bc{1} 0.62   & \bc{0} 0.00   & \gc{0} 0.00   & \gc{1} 0.83   & \gc{0} 0.00   & \gc{2} 1.67   & \pc{1} 0.63 \\
\bottomrule
\end{tabular}%
}
\end{table*}

\begin{table*}[t]
\centering
\caption{Ablation experiments on the number of search. In the Model column, the first number in parentheses indicates the maximum number of search for image/text search tasks, and the second number indicates the maximum number of search for video search tasks.}
\label{tab:retry_ablation}
\setlength{\tabcolsep}{6pt} 
\renewcommand{\arraystretch}{1.3} 
\resizebox{0.81\textwidth}{!}{
\begin{tabular}{l 
                *{4}{>{\centering\arraybackslash}p{1.3cm}} 
                *{4}{>{\centering\arraybackslash}p{1.3cm}} 
                >{\centering\arraybackslash}p{1.3cm}}
\toprule
\multirow{2}{*}[-0.5ex]{\textbf{Model}} 
& \multicolumn{4}{c}{\textbf{Retrieval Target Modality}} 
& \multicolumn{4}{c}{\textbf{Audio Content Type}} 
& \multirow{2}{*}{\textbf{Avg}} \\
\cmidrule(lr){2-5} \cmidrule(lr){6-9} 
 & SINGLE & MULTI & IMAGE & VIDEO & SPEECH & MUSIC & BIO & ENV &  \\
\midrule
\addlinespace[2pt]
Gemini-3-Pro (5,1)  & 43.75 & 22.50 & 31.25 & 18.75 & 29.17 & 29.17 & 41.67 & 25.00 & 29.06 \\
\addlinespace[2pt]
\hdashline
\addlinespace[2pt]
Gemini-3-Pro (10,3) & 57.50 & 40.63 & 38.75 & 36.88 & 55.00 & 46.67 & 39.17 & 36.67 & 43.44 \\
\addlinespace[2pt]
\hdashline
\addlinespace[2pt]
Gemini-3-Pro (15,5) & 56.25 & 38.75 & 50.00 & 31.25 & 70.83 & 50.00 & 41.67 & 20.83 & 44.06 \\
\bottomrule
\end{tabular}%
}
\end{table*}

\begin{table*}[t]
\centering
\caption{Ablation experiments on audio entities. ``Inferencing Audio Entities'' indicates the model's accuracy when only required to output the correct entity, and ``Providing Audio Entity Search Answers'' indicates the model's accuracy when the entity is provided to the model.}
\label{tab:entity_ablation}
\setlength{\tabcolsep}{4pt} 
\renewcommand{\arraystretch}{1.3} 
\resizebox{0.77\textwidth}{!}{
\begin{tabular}{l 
                *{4}{>{\centering\arraybackslash}p{1.3cm}} 
                *{4}{>{\centering\arraybackslash}p{1.3cm}} 
                >{\centering\arraybackslash}p{1.3cm}} 
\toprule
\multirow{2}{*}[-0.5ex]{\textbf{Model}} 
& \multicolumn{4}{c}{\textbf{Retrieval Target Modality}} 
& \multicolumn{4}{c}{\textbf{Audio Content Type}} 
& \multirow{2}{*}{\textbf{Avg}} \\
\cmidrule(lr){2-5} \cmidrule(lr){6-9} 
 & SINGLE & MULTI & IMAGE & VIDEO & SPEECH & MUSIC & BIO & ENV &  \\
\midrule
\multicolumn{10}{l}{\textbf{Inferencing Audio Entities}} \\
\hdashline
\addlinespace[2pt]
Gemini-3-Pro    & 40.63 & 19.38 & 34.38 & 40.63 & 75.00 & 12.50 & 37.50 & 29.17 & 33.76 \\
\addlinespace[1pt]
Mimo-V2.5       & 15.63 & 0.00  & 15.63 & 18.75 & 29.17 & 0.00  & 29.17 & 8.33  & 12.50 \\
\midrule
\multicolumn{10}{l}{\textbf{Providing Audio Entity Search Answers}} \\
\hdashline
\addlinespace[2pt]
Gemini-3-Pro    & 62.50 & 43.75 & 53.13 & 40.63 & 66.67 & 62.50 & 54.17 & 25.00 & 50.00 \\
\addlinespace[1pt]
Mimo-V2.5       & 21.88 & 13.13 & 34.38 & 18.75 & 29.17 & 29.17 & 29.17 & 12.50 & 22.03 \\
\midrule
\multicolumn{10}{l}{\textbf{End-to-End Omni-DeepSearch}} \\
\hdashline
\addlinespace[2pt]
Gemini-3-Pro                  &  57.50 &  40.63 &  38.75 & 36.88 &  55.00 &  46.67 & 39.17 &  36.67 &  {43.44} \\
\addlinespace[1pt]
Mimo-V2.5                    &  15.00 &  9.38  &  14.38 &  8.13  &  15.83 &  6.67  &  15.83 &  11.67 &  11.72 \\
\bottomrule
\end{tabular}%
}
\end{table*}

\textbf{Non-linguistic acoustic signals remain a major bottleneck.}
Across audio content types, speech is generally easier than non-speech audio. Gemini-3-Pro achieves 55.00\% accuracy on speech, compared with 39.17\% on animal sound and 36.67\% on ambient sound. This gap suggests that models are still better at exploiting linguistic and speaker-related cues than at interpreting non-linguistic acoustic signals. Music also remains challenging, as successful reasoning often requires recognizing melodies, instruments, or cultural references and linking them to external evidence.

\textbf{Open-source models still lag significantly behind in audio-driven deep search.}
There is a clear gap between closed-source and open-source models. While Gemini-3-Pro reaches 43.44\% average accuracy, the best open-source model, Mimo-V2.5, achieves only 11.72\%. Qwen3-Omni-30B-A3B (Thinking) outperforms its instruct variant, indicating that explicit reasoning behavior is beneficial for audio-driven deep search. Nevertheless, all open-source models remain limited on this benchmark, especially on video search and multi-audio text search, highlighting the difficulty of combining audio perception, tool use, and cross-modal multi-hop reasoning.

\subsection{Ablation Study}

\textbf{Increasing the search budget helps, but the gains saturate.}
We study the effect of search budget using Gemini-3-Pro. In Table~\ref{tab:retry_ablation}, the two numbers in parentheses denote the maximum retries for text/image-text search tasks and video search tasks, respectively. Increasing the budget from (5,1) to (10,3) improves average accuracy from 29.06\% to 43.44\%, showing the importance of iterative retrieval. Further increasing the budget to (15,5) brings only a small gain to 44.06\%, suggesting that the main bottlenecks also lie in audio entity inference, query formulation, and cross-modal verification. Larger search budgets may also introduce retrieval noise, as shown in \ref{3}.

\textbf{Audio entity inference and downstream search exhibit a synergistic effect for stronger models.}
Table~\ref{tab:entity_ablation} shows that Gemini-3-Pro achieves 33.76\% accuracy when directly identifying audio entities, but reaches 43.44\% in the end-to-end setting. This suggests that strong models can use question context, retrieval feedback, and intermediate evidence to refine audio entity inference during search. Providing the correct audio entity further improves Gemini-3-Pro to 50.00\%, confirming that audio entity inference remains important while downstream retrieval and verification are also challenging. In contrast, Mimo-V2.5 shows much weaker search-guided refinement, with 12.50\% entity identification accuracy and 11.72\% end-to-end accuracy.

\subsection{Case Study}

To better understand the challenges of Omni-DeepSearch, we analyze representative failure cases across task types and models, with details in Appendix~\ref{app:case_study}. The cases show that failures usually arise from the interaction of audio entity inference, search strategy, tool use, and cross-modal verification.

\textbf{Multi-audio tasks require balanced use of all clips.}
In multi-audio text search, a single misidentified clip can break the inference of the shared mediator entity. Models also tend to follow the clearest audio clip as the dominant clue, treating other clips as weak evidence rather than parallel constraints. This leads the search away from the true intersection among all audio inputs. Examples are shown in Appendix~\ref{app:case_study_multi_audio}.

\textbf{Image-text tasks require both retrieval and visual verification.}
Image-text failures mainly come from failing to retrieve the correct image, misreading fine-grained details in the correct image, or falling back to text search when image search fails. These cases show that visual retrieval quality and reliable visual inspection are both necessary. Examples are shown in Appendix~\ref{app:case_study_image_text}.

\textbf{Some failures are model-specific.}
Mimo-V2.5 struggles with music tasks involving niche genres, instruments, or performers, where subtle acoustic patterns must be mapped to real-world entities. Qwen3-Omni-30B-A3B (Thinking) often generates overly specific video queries with uncertain visual details while omitting the key audio-related entity. Examples are shown in Appendix~\ref{app:case_study_music_video}.

\textbf{Weaker models often fail before reasoning.}
For weaker models, failures often occur at the tool-use level. Gemini-2.5-Flash-Lite may generate empty tool calls despite expressing valid search intentions, while Qwen2.5-Omni-3B may produce malformed tool calls, repetitive outputs, or abandon the task after failed searches. Examples are shown in Appendix~\ref{app:case_study_tool_use}.

\textbf{Speech can still be difficult.}
Although speech is easier for strong models, weaker models may confuse speakers with similar timbre, accent, or speaking style. They may also over-rely on spoken content and ignore acoustic identity cues when the narrative matches a plausible but incorrect entity. Examples are shown in Appendix~\ref{app:case_study_speech}.



\section{Conclusion}

We introduced Omni-DeepSearch, a benchmark for audio-driven omni-modal deep search. Starting from audio alone, models must retrieve and reason over text, image, and video evidence. Experiments on 640 samples across 15 categories show that existing models remain limited, with the strongest model reaching only 43.44\% average accuracy. Our analyses identify audio entity inference, query formulation, tool-use reliability, and cross-modal verification as key bottlenecks, highlighting audio-driven deep search as a challenging direction for future multimodal agents.

\setcitestyle{numbers,square}
\bibliography{citation}

\newpage

\newpage
\EnableTOC
\clearpage
\appendix

\section*{Appendix}
\begingroup
\setcounter{tocdepth}{2}  
\tableofcontents
\endgroup

\section{Case Study}
\label{app:case_study}

\subsection{Multi-Audio Failure Cases}
\label{app:case_study_multi_audio}

\begin{tcolorbox}[
    title=\textbf{Case Study: Multi-Audio Reasoning Failure},
    fonttitle=\bfseries,
    breakable,
    fontupper=\small,
    colback=gray!10,    
    colframe=black,    
    coltitle=white,
    colbacktitle=black
]
\noindent
\begin{tabularx}{\textwidth}{@{}X m{5.5cm}@{}}
    \textbf{Question:} A specific feature film serves as the unique intersection of the provided audio tracks. One of the production companies behind this film's broader franchise was founded by a specific duo. In late 2025, this production company hired a seasoned producer and executive to lead its film division. This executive produced a feature film that incorporates a specific regional criminal coalition as a plot element. A prominent historical leader of this coalition was eventually found murdered in a specific seaside resort town. What is the name of this town? 
    & 
    \centering
    \begin{tcolorbox}[
        colback=white, colframe=black, width=4.8cm, arc=2pt, boxrule=0.5pt,
        left=2pt, right=2pt, top=2pt, bottom=2pt, boxsep=0pt, halign=center
    ]
        \includegraphics[width=4.4cm]{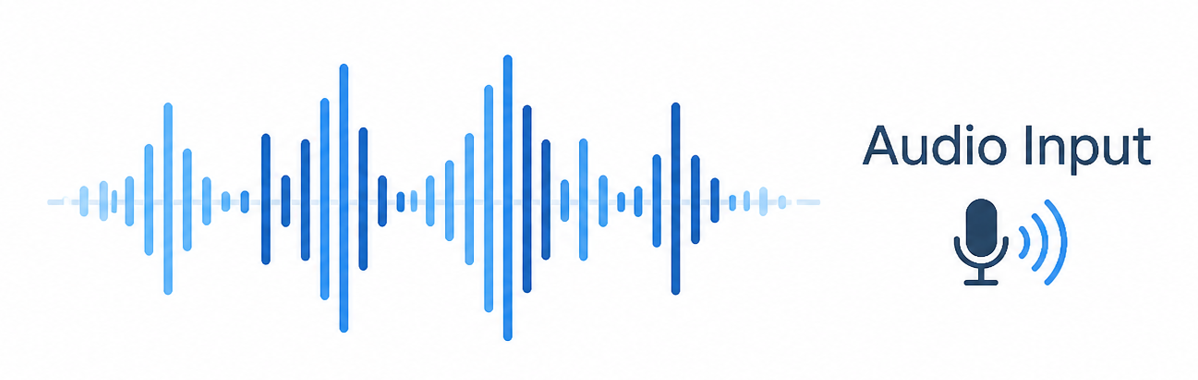}
    \end{tcolorbox}
    \vspace{2pt}
    \flushleft \scriptsize
    The input consists of four synchronized tracks: the voice of \textbf{Alfie Allen} (\textbf{Speech}), the song ``Killing Strangers'' by \textbf{Marilyn Manson} (\textbf{Music}), a \textbf{Beagle} barking (\textbf{Bio}), and the engine of a \textbf{Ford Mustang Mach 1} (\textbf{Env}).
\end{tabularx}

\vspace{0.8em}
\tcbline 
\vspace{0.8em}

\textbf{Model Reasoning: Turn 1 (Initial Misidentification)} \par
\vspace{0.5em}
\textbf{Thought:} Audio Subject: The intersection is ``The World's End'' film franchise (Cornetto Trilogy). The audio contains:
\begin{itemize}
    \item \textbf{Track 1 (Speech):} A man with a distinct British accent (likely Simon Pegg or Nick Frost).
    \item \textbf{Track 2 (Music):} A punk/industrial song with lyrics: ``The world will live in every house / It's hard to pick which one they hate the most / The horrible people, the horrible people...''
    \item \textbf{Track 3 (Bio):} A dog barking.
    \item \textbf{Track 4 (Env):} A high-performance car engine revving.
\end{itemize}
The song in Track 2 is ``The Horrible People'' by the band ``The Horrible People,'' which is featured in the film \textit{The World's End}. The speaker in Track 1 sounds like Simon Pegg. The car and dog are also elements found in the Cornetto Trilogy. The ``unique intersection'' is the film \textit{The World's End}.

\medskip
\textbf{Critical Error Analysis:} 
The model failed to recognize the actual speaker as \textbf{Alfie Allen} (from \textit{John Wick}) and misidentified the Marilyn Manson track as a fictional band. By mapping the speaker to Simon Pegg, it hallucinated a connection to the ``Cornetto Trilogy.'' This \textbf{audio-entity mismatch} caused the entire KG-path to deviate from the correct starting node (\textit{John Wick}).

\vspace{0.8em}
\tcbline
\vspace{0.8em}

\textbf{Outcome: Exhaustion of Search Budget} \par
\vspace{0.5em}
Due to the erroneous starting point, the agent entered a recursive failure loop. It spent all \textbf{10 allocated search turns} chasing false leads related to the wrong film franchise. For example:
\begin{itemize}
    \item \textbf{Turn 2:} Search for ``late 2025 hired seasoned producer executive lead film division production company Cornetto Trilogy'' yielded irrelevant Reddit leaks.
    \item \textbf{Turn 4:} After identifying Working Title Films (the wrong company), it searched for its 2025 executive appointments, leading to a dead end regarding the ``criminal coalition.''
\end{itemize}

\textbf{Final Result:} The model was unable to provide a valid answer after exhausting its search quota, as it remained trapped in the wrong knowledge domain throughout the process.
\end{tcolorbox}

\begin{tcolorbox}[
    title=\textbf{Case Study: Multi-Audio Reasoning Failure (Dominant Clue Bias)},
    fonttitle=\bfseries,
    breakable,
    fontupper=\small,
    colback=gray!10,    
    colframe=black,    
    coltitle=white,
    colbacktitle=black
]
\noindent
\begin{tabularx}{\textwidth}{@{}X m{5.5cm}@{}}
    \textbf{Question:} A specific feature film serves as the unique intersection of the provided audio tracks. This film includes a cast member whose sister is a professor at a specific university's business school. In 2025, this cast member served as a Fulbright Specialist at a specific Argentine university to help lay the groundwork for a unicorn-entrepreneur ecosystem. A former leader of this institution willed his personal property to the school, which later assumed formal control of the estate to operate it as a staff retreat and exhibition space. This estate is geographically situated within a specific city nestled in a valley formed by mountain ranges to its immediate east and west. What is the exact name of this city?
    & 
    \centering
    \begin{tcolorbox}[
        colback=white, colframe=black, width=4.8cm, arc=2pt, boxrule=0.5pt,
        left=2pt, right=2pt, top=2pt, bottom=2pt, boxsep=0pt, halign=center
    ]
        \includegraphics[width=4.4cm]{figs/audio_input.png}
    \end{tcolorbox}
    \vspace{2pt}
    \flushleft \scriptsize
    The input consists of three synchronized tracks: the voice of \textbf{Cillian Murphy} (\textbf{Speech}), a \textbf{Trombone-driven} orchestral score (\textbf{Music}), and the ambient interior noise of a \textbf{Boeing 747} (\textbf{Env}).
\end{tabularx}

\vspace{0.8em}
\tcbline 
\vspace{0.8em}

\textbf{Model Reasoning: Turn 1 (Heuristic Over-reliance)} \par
\vspace{0.5em}
\textbf{Thought:} Audio Subject: The intersection is the film \textbf{``Oppenheimer.''} 
The audio contains:
\begin{itemize}
    \item \textbf{Track 1 (Speech):} Identified as Cillian Murphy (Speech thanking Nolan/Thomas).
    \item \textbf{Track 2 (Music):} A generic jazz-influenced track likely from the film's score or promotional material.
    \item \textbf{Track 3 (Env):} Interpreted as the \textbf{thunderous blast of an atomic bomb} or the shockwave of a nuclear explosion.
\end{itemize}
The unique intersection is identified as \textit{Oppenheimer}. 

\medskip
\textbf{Sub-Question 1:} Identify the cast member of \textit{Oppenheimer} whose sister is a professor at a business school.

\vspace{0.8em}
\tcbline
\vspace{0.8em}

\textbf{Critical Failure Analysis: The Heuristic Override} \par
\vspace{0.5em}
The model exhibited a severe **Dominant Clue Bias**. Upon recognizing Cillian Murphy’s voice, it immediately defaulted to the film \textit{Oppenheimer}—his most prominent recent work. This caused the model to perform a ``best-fit'' interpretation of the other tracks rather than a logical intersection:
\begin{enumerate}
    \item \textbf{The Music:} It ignored the specific ``BRAAAM'' trombone score from Hans Zimmer's \textit{Inception}, labeling it as generic promotional background music.
    \item \textbf{The Environment:} It misidentified the specific \textbf{external roar of a Boeing 747} as an \textbf{atomic bomb explosion}. Although the sound is objectively the massive exterior noise of an aircraft (the primary setting of \textit{Inception}'s climax is aboard a flying Boeing 747), the model forced this auditory evidence to align with the nuclear detonation scene in \textit{Oppenheimer}.

\end{enumerate}
The true intersection is \textbf{\textit{Inception}}. By anchoring on the most famous association (Murphy $\rightarrow$ \textit{Oppenheimer}), the model’s reasoning chain drifted before the search even began.

\vspace{0.8em}
\tcbline
\vspace{0.8em}

\textbf{Outcome: Exhaustion of Search Budget} \par
\vspace{0.5em}
By locking onto the wrong knowledge domain, the agent spent all \textbf{10 allocated search turns} investigating the wrong actors. The search trajectory illustrates a recursive failure:
\begin{itemize}
    \item \textbf{Turns 1--5:} It searched for ``Cillian Murphy's sister professor business school,'' finding irrelevant biographical snippets.
    \item \textbf{Turns 6--10:} It expanded to other \textit{Oppenheimer} cast members (Emily Blunt, Robert Downey Jr.), attempting to force a connection between their relatives and the Fulbright program in Argentina.
\end{itemize}

\textbf{Conclusion:} 
The agent **exhausted its search quota** without ever questioning its initial film identification. This case demonstrates that when a model treats one modality as a "dominant signal" rather than an equal constraint, it will actively misinterpret environmental and musical cues to fit a flawed hypothesis, derailing the entire multi-hop reasoning path.
\end{tcolorbox}

\subsection{Image-Text Failure Cases}
\label{app:case_study_image_text}
\begin{tcolorbox}[
    title=\textbf{Case Study: Failing to Retrieve the Correct Image},
    fonttitle=\bfseries,
    breakable,
    fontupper=\small,
    colback=gray!10,    
    colframe=black,    
    coltitle=white,
    colbacktitle=black
]
\noindent
\begin{tabularx}{\textwidth}{@{}X m{5.5cm}@{}}
    \textbf{Question:} Identify the type of machine operating in the audio. A specific English manufacturer of these machines was purchased by a larger locomotive company following the collapse of an agricultural combine in the 1930s. Trace this to the heritage site that now preserves part of this manufacturer's original works. In an exterior image of this site, what specific two-word name is painted on the side of the vintage machine parked directly in front of the white 'MUSEUM' entrance?
    & 
    \centering
    \begin{tcolorbox}[
        colback=white, colframe=black, width=4.8cm, arc=2pt, boxrule=0.5pt,
        left=2pt, right=2pt, top=2pt, bottom=2pt, boxsep=0pt, halign=center
    ]
        \includegraphics[width=4.4cm]{figs/audio_input.png}
    \end{tcolorbox}
    \vspace{2pt}
    \flushleft \scriptsize
    The input features the rhythmic mechanical sounds of a \textbf{J.I. Case steam tractor} (\textbf{Env}), characterized by heavy chuffing and metallic clanking.
\end{tabularx}

\vspace{0.8em}
\tcbline 
\vspace{0.8em}

\textbf{Model Reasoning: Turns 1--3 (Successful Multi-Hop Linking)} \par
\vspace{0.5em}
The agent effectively utilized text-based tools to navigate a complex historical trajectory:
\begin{itemize}
    \item \textbf{Initial Search:} It identified the machine as a steam tractor and traced the 1932 collapse of the \textbf{Agricultural \& General Engineers (AGE)} combine.
    \item \textbf{Entity Linking:} It successfully identified \textbf{Richard Garrett \& Sons} as the manufacturer and its subsequent acquisition by Beyer, Peacock \& Company.
    \item \textbf{Location Anchoring:} It correctly localized the surviving works to the \textbf{Long Shop Museum} in Leiston. 
\end{itemize}
By Turn 3, the agent had accurately identified the correct geographical and historical node.

\vspace{0.8em}
\tcbline
\vspace{0.8em}

\textbf{Critical Failure Analysis: Rigid Query Refinement} \par
\vspace{0.5em}
The failure occurred exclusively in the \textbf{visual retrieval phase} (Turns 4--10). Despite having the correct museum name, the agent failed to retrieve a high-quality image of the machine:
\begin{enumerate}
    \item \textbf{Query Monotony:} The model became trapped in a loop of near-identical queries (e.g., "Long Shop Museum white entrance building machine"). It focused on the same set of keywords for 7 consecutive turns, which only yielded generic exterior shots of the museum.
    \item \textbf{Retrieval Failure:} None of the retrieved images provided sufficient clarity to read the name on the machine's side. The model failed to pivot its strategy—such as searching for specific museum exhibits, "portable steam engine" archives, or higher-resolution Getty/Alamy stock photos of that specific site.
    \item \textbf{Missing Visual Signal:} Because the agent never successfully triggered the retrieval of the correct image (showing the text "QUEEN VICTORIA"), the reasoning chain was physically blocked by a lack of input data.
\end{enumerate}

\vspace{0.8em}
\tcbline
\vspace{0.8em}

\textbf{Outcome: Termination by Quota Exhaustion} \par
\vspace{0.5em}
Unlike cases of hallucination, this agent exhibited a \textbf{hard failure} due to resource limits:
\begin{itemize}
    \item \textbf{Exhaustion:} After 10 turns of micro-adjusting the same failed search query, the agent reached its maximum step limit.
    \item \textbf{Result:} It terminated without providing a final answer, as the "Visual Verification" step remained unfulfilled.
\end{itemize}

\textbf{Conclusion:} 
This case highlights a Search Strategy Stalemate. The model's inability to diversify its visual search queries prevented it from acquiring the necessary visual evidence. As a result, even with flawless text-based reasoning, the agent failed to arrive at the final answer: it never obtained the critical visual information.
\end{tcolorbox}

\begin{tcolorbox}[
    title=\textbf{Case Study: Misreading Fine-Grained Details in the Correct Image},
    fonttitle=\bfseries,
    breakable,
    fontupper=\small,
    colback=gray!10,    
    colframe=black,    
    coltitle=white,
    colbacktitle=black
]
\noindent
\begin{tabularx}{\textwidth}{@{}X m{5.5cm}@{}}
    \textbf{Question:} Identify the surname of the featured solo performer in this audio. A recurring television antagonist sharing this surname is canonically an alumnus of a specific private research university. Find the primary rare book facility within this university's library system. In a building image of this facility's exterior, how many horizontal rows of panels make up the elevated main block?
    & 
    \centering
    \begin{tcolorbox}[
        colback=white, colframe=black, width=4.8cm, arc=2pt, boxrule=0.5pt,
        left=2pt, right=2pt, top=2pt, bottom=2pt, boxsep=0pt, halign=center
    ]
        \includegraphics[width=4.4cm]{figs/audio_input.png}
    \end{tcolorbox}
    \vspace{2pt}
    \flushleft \scriptsize
    The audio features a solo horn performance. The soloist is \textbf{Eric Terwilliger} (noted for his work on major film scores).
\end{tabularx}

\vspace{0.8em}
\tcbline 
\vspace{0.8em}

\textbf{Model Reasoning: Turns 1--7 (Knowledge Node Localization)} \par
\vspace{0.5em}
The agent demonstrates robust textual reasoning to bridge the multi-hop gap:
\begin{itemize}
    \item \textbf{Entity Connection:} It links the soloist’s surname (\textbf{Terwilliger}) to the antagonist \textbf{Robert Terwilliger} (\textit{Sideshow Bob}).
    \item \textbf{Academic Link:} It correctly identifies Sideshow Bob as a \textbf{Yale University} alumnus.
    \item \textbf{Facility Search:} It locates the \textbf{Beinecke Rare Book \& Manuscript Library} as the target facility.
    \item \textbf{Successful Retrieval:} In Turn 7, the model uses the tool \texttt{image\_search\_by\_text\_query} to retrieve a high-quality, clear exterior image of the library.
\end{itemize}

\vspace{0.8em}
\tcbline
\vspace{0.8em}

\textbf{Critical Failure Analysis: Flawed Visual Segmentation} \par
\vspace{0.5em}
The failure in the final step reveals a significant deficit in the model's **multi-modal spatial reasoning**, where the visual analysis itself was fundamentally problematic:
\begin{enumerate}
    \item \textbf{Spatial Miscounting:} Despite having a clear view of the facade, the agent failed to accurately segment the horizontal grid. It incorrectly mapped the physical boundaries of the marble panels, likely mistaking the structural granite ribs or the roof-line transition as an additional row.
    \item \textbf{Visual-Textual Confirmation Bias:} The model’s visual analysis was lazy; it attempted to "see" what it had already "read" in the text (the "six-story" stack description). Instead of performing an objective count of the 5 visible rows, it hallucinated a 6th row to align the image with its textual belief.
    \item \textbf{Lack of Self-Correction:} The model explicitly listed "Row 1 to Row 6" in its thoughts, indicating that its visual processing unit is unable to provide a high-fidelity "count-and-verify" signal that can override flawed internal hypotheses.
\end{enumerate}

\vspace{0.8em}
\tcbline
\vspace{0.8em}

\textbf{Outcome: Final Answer Error} \par
\vspace{0.5em}
The agent processed the correct visual evidence but reached an incorrect conclusion due to imprecise spatial parsing.
\begin{itemize}
    \item \textbf{Agent Answer:} 6
    \item \textbf{Ground Truth:} 5
\end{itemize}

\textbf{Conclusion:} 
This case demonstrates that **visual retrieval does not guarantee visual understanding**. The agent’s spatial reasoning was insufficient to distinguish between the 5 actual architectural rows and its expected count of 6, highlighting a failure where the visual modality acts as a "rubber stamp" for textual bias rather than an independent source of truth.
\end{tcolorbox}

\begin{tcolorbox}[
    title=\textbf{Case Study: Falling Back to Text Search When Image Search Fails},
    fonttitle=\bfseries,
    breakable,
    fontupper=\small,
    colback=gray!10,    
    colframe=black,    
    coltitle=white,
    colbacktitle=black
]
\noindent
\begin{tabularx}{\textwidth}{@{}X m{5.5cm}@{}}
    \textbf{Question:} Identify the device heard in the audio. The historical predecessor to this tool featured a puncturing mechanism inspired by a specific earlier communication apparatus. In an image of this apparatus, how many circular cutouts are visible on the outer face of the wheel holding the paper tape?
    & 
    \centering
    \begin{tcolorbox}[
        colback=white, colframe=black, width=4.8cm, arc=2pt, boxrule=0.5pt,
        left=2pt, right=2pt, top=2pt, bottom=2pt, boxsep=0pt, halign=center
    ]
        \includegraphics[width=4.4cm]{figs/audio_input.png}
    \end{tcolorbox}
    \vspace{2pt}
    \flushleft \scriptsize
    Audio Input: Buzzing sound of a coil machine, identified as a \textbf{Electric Tattoo Machine}.
\end{tabularx}

\vspace{0.8em}
\tcbline 
\vspace{0.8em}

\textbf{Model Reasoning: Tracing to the Automatic Telegraph} \par
\vspace{0.5em}
The agent demonstrates high accuracy in identifying the historical chain:
\begin{itemize}
    \item \textbf{Historical Link:} It correctly identifies the tattoo machine's ancestor as **Thomas Edison's Electric Pen** (1875).
    \item \textbf{Inspiration Source:} It traces the pen's puncturing mechanism to the **Edison Automatic Telegraph** perforator.
\end{itemize}

\vspace{0.8em}
\tcbline
\vspace{0.8em}

\textbf{Critical Failure Analysis: Strategic Pivot to Textual Evidence} \par
\vspace{0.5em}
The model's primary failure is its decision to abandon visual observation for textual confirmation:
\begin{enumerate}
    \item \textbf{Visual Skepticism :} After retrieving images of the Edison Telegraph (Turn 6), the agent expresses uncertainty in its thought log, perceiving that the images lack the clarity or specific "angle" needed to definitively count the holes. 
    \item \textbf{The Textual Shift:} Instead of searching for higher-resolution visual close-ups, the agent makes a strategic pivot. It assumes the count must be explicitly mentioned in historical documentation. It spends the remaining turns scouring patent texts and museum records for a phrase like "wheel with 8 holes."
    \item \textbf{Information Stalemate:} Because historical patent descriptions often detail the *function* of a gear rather than its aesthetic cutout count, the model finds zero textual results. This results in a "search spiral" where the model repeatedly micro-adjusts its textual queries, hoping to find a written confirmation that does not exist.
\end{enumerate}

\vspace{0.8em}
\tcbline
\vspace{0.8em}

\textbf{Outcome: Final Turn Budget Exhaustion} \par
\vspace{0.5em}
The agent's refusal to rely on the visual modality leads to a hard failure.
\begin{itemize}
    \item \textbf{Final Result:} By Turn 10, the agent is still requesting patent text for "paper tape wheel holes count." It exhausts its budget and terminates by outputting a JSON tool-call instead of a verified numerical answer.
\end{itemize}

\textbf{Conclusion:} 
This case highlights a **Multi-modal Confidence Gap**. The agent did not fail due to an incorrect reasoning chain, but because it **distrusted its ability to extract facts from images**. By pivoting to a text-only search for a visual-only attribute, the model effectively disabled its own visual capabilities, leading to a dead-end search for a textual fact that was never recorded in written history.
\end{tcolorbox}

\subsection{Music and Video Failure Cases}
\label{app:case_study_music_video}

\begin{tcolorbox}[
    title=\textbf{Case Study: Niche Audio Misclassification },
    fonttitle=\bfseries,
    breakable,
    fontupper=\small,
    colback=gray!10,    
    colframe=black,    
    coltitle=white,
    colbacktitle=black
]
\noindent
\begin{tabularx}{\textwidth}{@{}X m{5.5cm}@{}}
    \textbf{Question:} The ensemble performing this traditional instrumental excerpt shares its exact name with a village located in a specific nation. In 2026, this ensemble was scheduled to perform at a specific venue. Excluding its definite article, reversing the two words of this venue's name yields the exact name of a specific unincorporated community. This community is situated along a specific east-west state highway. As this highway continues east, it intersects a village that features a central square previously used as a college campus, and which later underwent multiple district reorganizations due to the loss of separate school buildings to fires. What is the name of this village?
    & 
    \centering
    \begin{tcolorbox}[
        colback=white, colframe=black, width=4.8cm, arc=2pt, boxrule=0.5pt,
        left=2pt, right=2pt, top=2pt, bottom=2pt, boxsep=0pt, halign=center
    ]
        \includegraphics[width=4.4cm]{figs/audio_input.png}
    \end{tcolorbox}
    \vspace{2pt}
    \flushleft \scriptsize
    The audio features a traditional instrumental piece by the Tuvan ensemble \textbf{Alash} (named after the Alash village in Tuva, Russia).
\end{tabularx}

\vspace{0.8em}
\tcbline 
\vspace{0.8em}

\textbf{Model Reasoning: Turns 1--4 (Category Misidentification)} \par
\vspace{0.5em}
The agent fails at the very first step of multi-modal integration—audio classification:
\begin{itemize}
    \item \textbf{Turn 1-2:} Upon hearing the flute and string textures, the model defaults to a high-probability heuristic: \textbf{Irish/Celtic folk music}. It searches for "Irish traditional ensemble named after a village" and identifies groups like \textit{Lúnasa} and \textit{Villagers}.
    \item \textbf{Turn 3-4:} The model anchors onto the \textbf{Tulla Céilí Band} (named after Tulla, Ireland). It finds a valid 2026 performance schedule for this band at "Kennedy Hall" and "The London Irish Centre."
    \item \textbf{The Anchor Bias:} Because the Tulla Céilí Band fits the "named after a village" and "2026 performance" constraints, the model commits to this path, ignoring the actual sonic characteristics of Tuvan music (throat singing overtones and specific string timbres).
\end{itemize}

\vspace{0.8em}
\tcbline
\vspace{0.8em}

\textbf{Critical Failure Analysis: Popularity Bias \& Logical Dead-end} \par
\vspace{0.5em}
The model's failure stems from an initial misclassification that propagated through the entire reasoning chain, driven by two primary factors:
\begin{enumerate}
    \item \textbf{Fame Heuristic Over Precision:} The model exhibited a strong preference for "famous" entities over "niche" ones. Upon recognizing "traditional instrumental music" and "village name," it bypassed the specific auditory signatures of Tuvan throat singing (the ensemble \textbf{Alash}) and defaulted to the high-frequency category of \textbf{Irish Folk}. By choosing the well-known \textbf{Tulla Céilí Band} over the correct but less prominent \textbf{Alash}, the model prioritized a "popular guess" that doomed the subsequent geographic search.
    \item \textbf{Recursive Logical Dead-end:} This initial "Popularity Bias" led to a structural failure in the "Name Reversal" step. Since the model was anchored to the Irish ensemble, it attempted to force the logic on the venue "**Kennedy Hall**." Its search for the non-existent unincorporated community "**Hall Kennedy**" on a state highway created a recursive loop. The model failed to realize that the true starting point (\textbf{Alash}) would have led to the venue "**Old Town School of Folk Music**," which perfectly satisfies the "Town Old" (MD) reversal and the subsequent path to **Mount Morris, Illinois**.
\end{enumerate}

\vspace{0.8em}
\tcbline
\vspace{0.8em}

\textbf{Outcome: Search Quota Exhaustion} \par
\vspace{0.5em}
The agent exhausted its \textbf{10-turn search budget} by investigating irrelevant Irish cultural geography.
\begin{itemize}
    \item \textbf{Final Result:} Failure. The model remained trapped in a "Popularity Loop," attempting to verify Irish venues and highways in the United States, and never questioned its initial misidentification of the ensemble.
\end{itemize}

\textbf{Conclusion:} 
This case highlights the **Popularity Heuristic Trap** in multi-modal reasoning. When faced with ambiguous or cold-start audio inputs, the model tends to map them to the most "famous" or "frequently occurring" concepts in its training data (e.g., Irish Folk) rather than performing a precise zero-shot identification. This bias effectively "blinds" the model to the correct reasoning path, as it forces the remaining logical constraints to fit a fundamentally flawed initial premise.
\end{tcolorbox}

\begin{tcolorbox}[
    title=\textbf{Case Study: Search Query Over-Specification},
    fonttitle=\bfseries,
    breakable,
    fontupper=\small,
    colback=gray!10,    
    colframe=black,    
    coltitle=white,
    colbacktitle=black
]
\noindent
\begin{tabularx}{\textwidth}{@{}X m{5.5cm}@{}}
    \textbf{Question:} Identify the video using the provided audio, which captures a live set by a celebrated musician at a renowned North American music venue. In the video, the central performer is seen playing a stringed instrument and singing on stage. As the camera changes angles to provide a clear view of the performer's thick instrument strap, what specific shapes or symbols are prominently decorating it?
    & 
    \centering
    \begin{tcolorbox}[
        colback=white, colframe=black, width=4.8cm, arc=2pt, boxrule=0.5pt,
        left=2pt, right=2pt, top=2pt, bottom=2pt, boxsep=0pt, halign=center
    ]
        \includegraphics[width=4.4cm]{figs/audio_input.png}
    \end{tcolorbox}
    \vspace{2pt}
    \flushleft \scriptsize
    Audio Input: Performer: \textbf{Stevie Ray Vaughan}, Piece: \textbf{"Texas Flood"}, Instrument: \textbf{Electric Guitar} (featuring heavy wah-pedal effects).
\end{tabularx}

\vspace{0.8em}
\tcbline 
\vspace{0.8em}

\textbf{Model Reasoning: Turns 1--3 (Descriptive Search Loop)} \par
\vspace{0.5em}
The agent attempts to identify the video by translating its auditory observations into literal search queries:
\begin{itemize}
    \item \textbf{Turn 1:} The model correctly identifies the musical style (blues-rock) and the technical effect (\textbf{wah-pedal}). However, its search query is a mere concatenation of the question's descriptors: \textit{"live electric guitar solo wah pedal effects decorated instrument strap rock musician North American venue"}.
    \item \textbf{Turn 2:} After the first failure, it adds specific venue guesses (Madison Square Garden, Red Rocks) but maintains the same descriptive clutter, leading to another null result.
    \item \textbf{Turn 3:} The model correctly identifies that the sound is characteristic of artists like Jimi Hendrix or Stevie Ray Vaughan, yet it fails to search for these specific individuals, opting instead for generic terms like \textit{"iconic rock guitarist"}.
\end{itemize}

\vspace{0.8em}
\tcbline
\vspace{0.8em}

\textbf{Critical Failure Analysis: Lack of Entity-Centric Search} \par
\vspace{0.5em}
The model's reasoning exhibits a fundamental disconnect between perception and retrieval strategy:
\begin{enumerate}
    \item \textbf{Descriptive Overload:} Instead of identifying the \textbf{Subject} (the musician) first, the model treated the search engine like a visual captioning tool. Queries like "decorated instrument strap symbols" are too fine-grained and semantically noisy for broad video search engines, which prioritize entities (names, titles) over visual scene descriptions.
    \item \textbf{Failure to Anchor on the Audio:} The audio is the primary "anchor." A more effective strategy would have been to identify the specific song or performance via the unique guitar solo signature. By ignoring the "who" and focusing on the "what" (the strap), the model drifted into a combinatorial explosion of irrelevant search results.
    \item \textbf{Visual Hallucination (Final Turn):} In its desperation, the model began hallucinating specific symbols like "skull and crossbones" to narrow the search, moving further away from the ground truth (\textbf{Musical notes}).
\end{enumerate}

\vspace{0.8em}
\tcbline
\vspace{0.8em}

\textbf{Outcome: Recursive Failure \& Timeout} \par
\vspace{0.5em}
\begin{itemize}
    \item \textbf{Result:} Failure. The model spent its entire reasoning budget micro-adjusting descriptive queries without ever successfully identifying the performer or the specific concert video.
\end{itemize}

\textbf{Conclusion:} 
This case demonstrates a **Retrieval Strategy Deficiency**. The model failed to translate auditory "style" into a concrete "entity." By relying on literal, descriptive search terms rather than identifying the core subject (the musician), it created a "keyword soup" that made retrieval impossible. This highlights the need for models to prioritize entity-linking over scene-description when performing multi-modal tracing tasks.
\end{tcolorbox}

\subsection{Tool-Use Failure Cases}
\label{app:case_study_tool_use}
\begin{tcolorbox}[
    title=\textbf{Case Study: Tool Invocation Paralysis \& Systemic Execution Failure},
    fonttitle=\bfseries,
    breakable,
    fontupper=\small,
    colback=gray!10,    
    colframe=black,    
    coltitle=white,
    colbacktitle=black
]
\noindent
\begin{tabularx}{\textwidth}{@{}X m{5.5cm}@{}}
    \textbf{Question:} Listen to the audio to identify the primary subject. This subject belongs to a specific genus. Another member of this genus was the focus of a threatened species assessment co-authored by researchers Urbani, Boubli, and Cortes-Ortíz. This assessed member is sometimes classified as a subspecies of a broader relative. A distinct population of this broader relative, found within a specific subnational department, was historically reclassified as its own separate entity. This department is the largest of its nation's nine constituent territories. In 2026, an airport in this region was heavily guarded by authorities during the transfer of a specific high-profile foreign fugitive. What is the name of this individual?
    & 
    \centering
    \begin{tcolorbox}[
        colback=white, colframe=black, width=4.8cm, arc=2pt, boxrule=0.5pt,
        left=2pt, right=2pt, top=2pt, bottom=2pt, boxsep=0pt, halign=center
    ]
        \includegraphics[width=4.4cm]{figs/audio_input.png}
    \end{tcolorbox}
    \vspace{2pt}
    \flushleft \scriptsize
    Audio Input: Deep, guttural primate vocalizations. Subject: \textbf{\textit{Alouatta caraya}} (Black Howler Monkey).
\end{tabularx}

\vspace{0.8em}
\tcbline 
\vspace{0.8em}

\textbf{Model Reasoning: Turns 1--5 (The "Ghost" Search Loop)} \par
\vspace{0.5em}
The agent shows perfect conceptual understanding but zero operational output:
\begin{itemize}
    \item \textbf{High-Fidelity Thought:} In Turn 1, the model correctly identifies the vocalizations as New World monkeys (Howler monkeys) and recognizes the significance of researchers \textbf{Urbani, Boubli, and Cortes-Ortíz}.
    \item \textbf{Structural Failure:} Despite formulating clear sub-questions in its mind (e.g., "What genus of primate is associated with Urbani?"), the actual tool call generated for \texttt{text\_search} contains an **empty query list** (\texttt{queries: []}).
    \item \textbf{Ineffective Self-Correction:} From Turn 2 to Turn 5, the model repeatedly acknowledges that "the previous search query was invalid" or that "the search tool continues to fail," yet it continues to output empty query lists in every subsequent attempt.
\end{itemize}

\vspace{0.8em}
\tcbline
\vspace{0.8em}

\textbf{Critical Failure Analysis: Silent Tool Call Paralysis} \par
\vspace{0.5em}
This case illustrates a unique failure mode where the model's action-generation logic is disconnected from its reasoning logic:
\begin{enumerate}
    \item \textbf{Format without Content:} The model successfully maintains the JSON/tool-calling format but fails to populate the specific arguments. This suggests a systemic breakdown in the final stage of response generation where the "intent" is lost before it reaches the "parameter" field.
    \item \textbf{The Dead-End Loop:} The agent becomes "aware" of the tool failure but is trapped in a deterministic cycle. It correctly reasons that it needs to search for the genus, yet every time it tries to act, the output pipeline results in a null query.
    \item \textbf{Cognitive Resignation:} By Turn 6, the model undergoes a "logical collapse." It concludes that since the tool is broken, the task is impossible, and explicitly gives up: \textit{"Since I cannot execute any searches... I cannot proceed with the task."}
\end{enumerate}

\vspace{0.8em}
\tcbline
\vspace{0.8em}

\textbf{Outcome: Premature Task Abandonment} \par
\vspace{0.5em}
\begin{itemize}
    \item \textbf{Result:} Failure. The model correctly identified the starting node (\textit{Alouatta}) and the middle nodes (Santa Cruz department in Bolivia, 2026 fugitive transfer) in its internal knowledge but could not verify them or bridge to the final answer (\textbf{Sebastian Marset}).
\end{itemize}

\textbf{Conclusion:} 
This failure represents an **Operational Execution Breakdown**. Unlike knowledge gaps or reasoning errors, this is a "tool paralysis" where the model understands the world and the task but is mechanically unable to interact with external tools.
\end{tcolorbox}

\begin{tcolorbox}[
    title=\textbf{Case Study: Structural Syntax Failure \& Tool Call Crash},
    fonttitle=\bfseries,
    breakable,
    fontupper=\small,
    colback=gray!10,    
    colframe=black,    
    coltitle=white,
    colbacktitle=black
]
\noindent
\begin{tabularx}{\textwidth}{@{}X m{5.5cm}@{}}
    \textbf{Question:} Identify the singer in the audio. A specific percussionist joined this singer's early backing band to complete its four-person lineup. Prior to this collaboration, the percussionist was employed as an in-house musician for a regional weekend radio broadcast. That broadcast derived its title from a book written by a specific author, who later contributed a non-fiction volume to a multi-volume regional profile series. In 2025, it was announced that a specific individual will release an upcoming project associated with this entity in 2026. Name this individual.
    & 
    \centering
    \begin{tcolorbox}[
        colback=white, colframe=black, width=4.8cm, arc=2pt, boxrule=0.5pt,
        left=2pt, right=2pt, top=2pt, bottom=2pt, boxsep=0pt, halign=center
    ]
        \includegraphics[width=4.4cm]{figs/audio_input.png}
    \end{tcolorbox}
    \vspace{2pt}
    \flushleft \scriptsize
    Audio Input: A classic rock-and-roll vocal performance identified as \textbf{Elvis Presley}.
\end{tabularx}

\vspace{0.8em}
\tcbline 
\vspace{0.8em}

\textbf{Model Reasoning: Turn 1 (The Immediate Execution Collapse)} \par
\vspace{0.5em}
The agent attempts to initiate the multi-hop search but fails at the syntax level:
\begin{itemize}
    \item \textbf{Cognitive Understanding:} In its \texttt{<think>} block, the model correctly paraphrases the multi-hop constraints: Singer $\rightarrow$ Percussionist (\textbf{D. J. Fontana}) $\rightarrow$ Radio Broadcast (\textbf{Louisiana Hayride}) $\rightarrow$ Series (\textbf{American Folkways}) $\rightarrow$ Upcoming project.
    \item \textbf{Syntactic Malformation:} When generating the \texttt{<tool\_call>}, the model injects non-standard arguments, specifically a \texttt{"toolbench\_rapidapi\_key"} field, and fails to maintain proper JSON syntax.
    \item \textbf{Parser Error:} The system returned a hard error: \textit{"Tool execution crashed: Expecting ',' delimiter"}. This indicates that the model's output stream generated an unparseable character sequence at the end of the JSON object.
\end{itemize}

\vspace{0.8em}
\tcbline
\vspace{0.8em}

\textbf{Outcome: Immediate System Halt} \par
\vspace{0.5em}
\begin{itemize}
    \item \textbf{Result:} Failure. The agent provided no answer.
    \item \textbf{Root Cause:} A "Structural Syntax Error" in the tool-call generation phase, preventing any multi-modal information retrieval.
\end{itemize}

\textbf{Conclusion:} 
This case demonstrates that **Agentic Reliability is as crucial as Reasoning Depth**. Even if a model perfectly "understands" the audio and the query, a single character-level error in its tool-calling format can lead to total task failure. This emphasizes the necessity for robust instruction-following and error-handling mechanisms in multi-modal agent systems.
\end{tcolorbox}

\subsection{Speech Failure Cases}
\label{app:case_study_speech}

\begin{tcolorbox}[
    title=\textbf{Case Study: Auditory Profiling Failure (Speaker Misidentification)},
    fonttitle=\bfseries,
    breakable,
    fontupper=\small,
    colback=gray!10,    
    colframe=black,    
    coltitle=white,
    colbacktitle=black
]
\noindent
\begin{tabularx}{\textwidth}{@{}X m{5.5cm}@{}}
    \textbf{Question:} The speaker portrayed a specific historical individual in a television film. During their medical studies, this individual was noted for making frequent trips to a nearby women's college. This individual was born in a specific city. In 2026, this city was selected to host an inaugural landowner field day organized by a prominent state forestry advocacy group. A notable figure associated with this advocacy group was the son of a banker and politician who is colloquially known as the 'father' of a specific county he helped establish. What is the exact name of this banker and politician?
    & 
    \centering
    \begin{tcolorbox}[
        colback=white, colframe=black, width=4.8cm, arc=2pt, boxrule=0.5pt,
        left=2pt, right=2pt, top=2pt, bottom=2pt, boxsep=0pt, halign=center
    ]
        \includegraphics[width=4.4cm]{figs/audio_input.png}
    \end{tcolorbox}
    \vspace{2pt}
    \flushleft \scriptsize
    Audio Input: A deep, resonant male voice with a classical British accent reading Sonnet 130. Actual Speaker: \textbf{Alan Rickman}.
\end{tabularx}

\vspace{0.8em}
\tcbline 
\vspace{0.8em}

\textbf{Detailed Analysis: The Auditory Anchor Failure} \par
\vspace{0.5em}
The reasoning chain collapsed at the initial modality-processing step because the model could not precisely distinguish between two actors with highly similar vocal profiles:
\begin{enumerate}
    \item \textbf{Timbre Style Confusion:} In Turn 1, the model correctly identified the content (Shakespeare's Sonnet 130) but attributed the voice to \textbf{Ralph Fiennes}. While both Rickman and Fiennes share a "refined British delivery," the model failed to detect the unique "languid" baritone and specific drawl that define Alan Rickman’s vocal fingerprint, mistaking it for Fiennes' slightly more melodic and breathy timbre.
    \item \textbf{Search Direction Misalignment:} This misidentification derailed the search for the "historical individual." The model looked for medical students played by Ralph Fiennes, which yielded no results. Fiennes played T.E. Lawrence in a TV movie, but Lawrence does not fit the "medical student" or "women's college" biographical clues.
\end{enumerate}

\vspace{0.8em}
\tcbline 
\vspace{0.8em}

\textbf{Conclusion:} 
\textbf{Identifying a unique speaker based on timbre, accent, and speaking style remains exceptionally challenging for multi-modal models.} In this case, the failure to profile the speaker’s unique auditory "fingerprint" caused the agent to drift from the actual historical figure (Blalock) into a heuristic-based search, ultimately failing to satisfy the multi-modal reasoning constraints of the challenge.
\end{tcolorbox}

\begin{tcolorbox}[
    title=\textbf{Case Study: Semantic Content Bias},
    fonttitle=\bfseries,
    breakable,
    fontupper=\small,
    colback=gray!10,    
    colframe=black,    
    coltitle=white,
    colbacktitle=black
]
\noindent
\begin{tabularx}{\textwidth}{@{}X m{5.5cm}@{}}
    \textbf{Question:} Identify Speaker 1 (the first voice heard, who commends the audience member). According to unverified biographical claims, this speaker immigrated to live with an aunt and uncle in a specific community. This community was founded by a plantation owner whose grandfather resided in a specific historic house until his death. The property for this house was previously purchased by a specific lieutenant-general. This official was a direct descendant of a specific historical figure. In 2025, a specific curator organized an exhibition exploring this figure's relationship to abstraction. What is the name of this curator?
    & 
    \centering
    \begin{tcolorbox}[
        colback=white, colframe=black, width=4.8cm, arc=2pt, boxrule=0.5pt,
        left=2pt, right=2pt, top=2pt, bottom=2pt, boxsep=0pt, halign=center
    ]
        \includegraphics[width=4.4cm]{figs/audio_input.png}
    \end{tcolorbox}
    \vspace{2pt}
    \flushleft \scriptsize
    Audio Input: A voice with a unique accent discussing a health battle and cancer remission. Actual Speaker: \textbf{Tommy Wiseau}.
\end{tabularx}

\vspace{0.8em}
\tcbline 
\vspace{0.8em}

\textbf{Model Reasoning: Turn 1 (Content-Driven Anchoring)} \par
\vspace{0.5em}
The agent fails at the initial speaker identification stage due to overwhelming semantic clues:
\begin{itemize}
    \item \textbf{The Semantic Trap:} The audio captures a moment where the speaker mentions a fan's relative surviving a health battle (cancer). The model immediately associates this with \textbf{Hugh Jackman}, who is globally famous for his publicized battle with skin cancer.
    \item \textbf{Premature Commitment:} In Turn 1, the model states with high confidence: \textit{"The audio features the distinct voice of actor Hugh Jackman... where he discusses a woman surviving a health battle (cancer in remission)." } This commitment occurs despite Tommy Wiseau's highly idiosyncratic and non-Australian accent.
\end{itemize}

\vspace{0.8em}
\tcbline 
\vspace{0.8em}

\textbf{Critical Failure Analysis: Heuristic Over-reliance} \par
\vspace{0.5em}
The model's initial misidentification leads to a recursive failure in solving the complex biographical riddle:
\begin{enumerate}
    \item \textbf{Forced Logic Alignment:} Because the model is convinced the speaker is Jackman, it attempts to force the "unverified biographical claim" (immigrating to live with an aunt and uncle) onto him. When search results (Turn 2, 4) confirm Jackman was born in Australia and stayed with his father, the model begins to hallucinate or search for "fake" biographies of Jackman instead of questioning the speaker's identity.
    \item \textbf{Topic Over Voice:} The model prioritized the \textbf{topic of the speech} (cancer) over the \textbf{acoustic profile} (the actual voice). Tommy Wiseau's accent is famously enigmatic and non-native, while Hugh Jackman's is a clear Australian-English. The model allowed the high-frequency "Jackman-cancer" association to override its auditory sensors.
    \item \textbf{Complexity Collapse:} The actual path (\textbf{Wiseau} $\rightarrow$ \textbf{Chalmette, LA} $\rightarrow$ \textbf{Duke of Kent House} $\rightarrow$ \textbf{Louise Bourgeois}) was never explored. The model remained trapped in a "Popularity Loop," eventually exhausting its budget trying to link Jackman to the community of Fairmount (conflating him with James Dean).
\end{enumerate}

\vspace{0.8em}
\tcbline 
\vspace{0.8em}

\textbf{Outcome: Final Turn Stalemate} \par
\vspace{0.5em}
\begin{itemize}
    \item \textbf{Result:} Failure. The model ran out of turns while trying to reconcile Hugh Jackman's life with the historical details of a plantation owner and a 2025 exhibition.
\end{itemize}

\textbf{Conclusion:} 
This case demonstrates a **Semantic Content Heuristic**. When audio contains a highly specific "buzzword" (e.g., cancer remission) associated with a world-famous celebrity, models tend to ignore conflicting auditory evidence (accents, timbre) and double down on the popular association. This "completionist" tendency to fit every clue to a famous entity prevents the model from identifying the correct, albeit more obscure, subject.
\end{tcolorbox}

\subsection{Example of Retrieval Noise Introduced by Over-Searching}
\label{3}

\begin{tcolorbox}[
    title=\textbf{Case Study: Over-Searching Introduces Retrieval Noise, Causing Reasoning Collapse},
    fonttitle=\bfseries,
    breakable,
    fontupper=\small,
    colback=gray!10,    
    colframe=black,    
    coltitle=white,
    colbacktitle=black
]

\noindent
\begin{tabularx}{\textwidth}{@{}X m{5.5cm}@{}}
    \textbf{Question:} Listen to the audio to determine the identity of the primary subject. This subject is situated at a specific junction where three massive structural segments diverge. In 2025, a sudden eruption and lava surge from this volcano produced a dense ash plume and reached a historic camp located within a specific country. The leadership of this country later sought to negotiate a new bilateral treaty to replace a prior interim accord that had granted a foreign allied power temporary control over its currency, exports, and certain border territories. One of these temporarily controlled territories was a distinct plateau region characterized by its iron-rich red soil and thorn-bush. This plateau is geographically separated from adjacent valleys to its south by a specific mountain range. What is the exact name of this mountain range?
    & 
    \centering
    \begin{tcolorbox}[
        colback=white, colframe=black, width=4.8cm, arc=2pt, boxrule=0.5pt,
        left=2pt, right=2pt, top=2pt, bottom=2pt, boxsep=0pt, halign=center
    ]
        \includegraphics[width=4.4cm]{figs/audio_input.png}
    \end{tcolorbox}
    \vspace{2pt}
    \flushleft \scriptsize
    The audio contains low-frequency volcanic rumbling and wind-like environmental noise associated with the active volcano \textbf{Erta Ale}.
\end{tabularx}

\vspace{0.8em}
\tcbline
\vspace{0.8em}

\textbf{Model Reasoning: Early Successful Retrieval} \par
\vspace{0.5em}

The agent initially follows the correct reasoning trajectory:
\begin{itemize}
    \item \textbf{Turn 1:} Correctly identifies the volcano as \textbf{Erta Ale} in Ethiopia.
    \item \textbf{Turn 2--3:} Correctly links the historical context to the \textbf{Anglo-Ethiopian Agreement} and identifies the \textbf{Haud Plateau}.
    \item \textbf{Turn 3:} Successfully retrieves the key evidence stating that the Haud is separated from the Ain and Nugal valleys by the \textbf{Buurdhaab mountain range}.
\end{itemize}

At this stage, the reasoning chain is already sufficient to produce the correct answer.

\vspace{0.8em}
\tcbline
\vspace{0.8em}

\textbf{Critical Failure Analysis: Over-Search Drift} \par
\vspace{0.5em}

Instead of terminating after obtaining the correct evidence, the 15-5 configuration continued searching for additional confirmation. This caused the reasoning process to gradually drift away from the correct hypothesis:
\begin{enumerate}
    \item \textbf{Hypothesis Expansion:} The model repeatedly introduced new candidate mountain ranges such as the \textbf{Audo Range}, \textbf{Karkaar Mountains}, and \textbf{Golis Mountains}, despite already retrieving the correct entity.
    
    \item \textbf{Retrieval Noise Accumulation:} Additional search turns produced geographically related but irrelevant snippets involving Somalia, Ogaden, and surrounding plateaus. These semantically plausible distractors weakened the model's confidence in the original correct reasoning path.
    
    \item \textbf{Belief Instability:} As the retrieval trajectory grew longer, the model continuously revised its own intermediate conclusions. Instead of consolidating evidence around ``Buurdhaab,'' it repeatedly reopened previously solved sub-questions and entered a recursive search loop.
\end{enumerate}

Unlike typical hallucination failures, the model had already retrieved the correct answer but lost it due to excessive search depth.

\vspace{0.8em}
\tcbline
\vspace{0.8em}

\textbf{Outcome: Reasoning Collapse After Excessive Search} \par
\vspace{0.5em}

The shorter 10-3 configuration successfully stopped after retrieving the correct evidence and returned the correct answer:
\begin{itemize}
    \item \textbf{10-3 Answer:} Buurdhaab
\end{itemize}

However, the longer 15-5 configuration exhausted its search budget while continuously exploring alternative geographic hypotheses:
\begin{itemize}
    \item \textbf{15-5 Result:} Failed to produce a final grounded answer.
\end{itemize}

\textbf{Conclusion:}
This case demonstrates that increasing search depth does not necessarily improve reasoning performance. Once the correct evidence has already been retrieved, additional search steps may instead introduce distractor information, destabilize intermediate beliefs, and ultimately cause reasoning collapse. The failure here is not caused by insufficient retrieval, but by the inability to stop searching after reaching a correct high-confidence state.
\end{tcolorbox}

\section{Data Generation Prompt}
\subsection{Single-Audio Text Data Generation Prompt}
\label{single}
\begin{tcolorbox}[
    title=\textbf{Single-Audio Text Data Generation Prompt},
    fonttitle=\bfseries,
    breakable, 
    fontupper=\small
]
You are an "Audio Search Challenge" Designer. 

1. You are given the REAL IDENTITIES of the speakers to help you disambiguate, but you must NEVER use these names in the final question. \\
\textbf{REMEMBER:} The solver has ONLY the audio segment and your question. They do NOT see the Video Title, Description.

2. \textbf{NODE SEQUENCE:} A logical chain of entities. Use the provided Node Article Snippets to discover how each entity connects to the next.

3. \textbf{NODE DESCRIPTIONS:} Factual snippets for each node in the path.

\textbf{Your Goal:} \\
Construct a multi-hop reasoning question where the solver must:
\begin{itemize}
    \item[1.] Listen to the audio to identify the "Start Node" (Speaker/Entity).
    \item[2.] Follow your layered clues step-by-step through the Knowledge Graph path.
    \item[3.] Identify one specific attribute (a name, a precise location, a specific date, or a technical term) from the final node in the path. The question must be constructed so that this attribute value is the only possible and exact answer.
\end{itemize}

\textbf{THE CLOAKING PROTOCOL (Anti-Semantic Leakage):}
\begin{itemize}
    \item \textbf{NO SEARCHABLE FINGERPRINTS:} Strictly forbid any specific quantities, unique descriptors, or highly specific biographical/historical anomalies.
    \item \textbf{BEWARE OF "SEMANTIC LEAKAGE":} Simply replacing proper nouns with complicated synonyms is a FAILURE if the underlying event is unique. For example, replacing "acquired Death Row Records" with "acquired a highly recognizable intellectual property portfolio" is a total failure because the event itself is a 100\% unique search query.
    \item \textbf{BLIND SEARCH TEST:} If the question text alone allows a search engine to find the answer without hearing the audio, the challenge is a total failure.
\end{itemize}

\textbf{CRITICAL RESTRICTION \& THE "BLAND BRIDGE" RULE:} \\
The question must NOT contain ANY real-world names, project titles, descriptive nouns, or thematic clues. 
\begin{itemize}
    \item \textbf{BIOGRAPHICAL LEAKAGE DEFENSE (SPEECH SPECIFIC):} Do NOT use unique biographical anomalies (e.g., fleeing a specific revolution, buying a specific famous company, committing a specific crime). Use generic professional collaborations (e.g., "a director they worked with in the early 2000s", "a co-author of their second publication").
    \item \textbf{DO NOT} use any information from the video title or description.
    \item \textbf{USE ONLY} abstract, neutral placeholders: "the entity," "the primary subject," "the individual," "the project."
    \item \textbf{NO MID-POINT SHORTCUTS (STRICT SEQUENTIAL DEPENDENCY):} The descriptive clues provided in the question MUST NOT be searchable on their own. Rule of Thumb: Read your generated question. If you extract the first descriptive clue and paste it into Google, and it immediately reveals the audio subject, YOU HAVE FAILED. 
    \item \textbf{DO NOT} use years (e.g., "1997"), decades (e.g., "the 90s"), or specific durations (e.g., "5 years later").
\end{itemize}

\textbf{ALGEBRAIC DEPENDENCY (X + Y = Z):} \\
The question must function like an algebraic equation: \\
\textbf{Variable X:} The identity/entity identified ONLY from the audio. \\
\textbf{Generic Relationship Y:} A broad, multi-valued clue provided in the text. \\
\textbf{Answer Z:} The intersection of X and Y.

\begin{itemize}
    \item[1.] \textbf{MINIMAL SUFFICIENT SPECIFICITY (ZERO AMBIGUITY):} The generated question MUST be concise, BUT uniqueness is paramount. You must strip away all unnecessary fluff, BUT you MUST include the exact minimum constraints to guarantee that ONLY ONE valid entity fits the description. Never create a "fork in the road" by being too vague.
    \item[2.] \textbf{DYNAMIC CLOAKING (FIRST HOP vs. SUBSEQUENT HOPS):} 
    \begin{itemize}
        \item \textbf{FIRST HOP (Absolute Cloaking):} The immediate link leaving the Start Node (the audio subject) MUST be absolutely minimal and generic. It must rely entirely on the audio identification to be solved.
        \item \textbf{SUBSEQUENT HOPS (Relative Cloaking):} As you move deeper into the chain, keep descriptions generic UNLESS disambiguation is needed. If a bland term (e.g., "a specific smaller community") could apply to multiple valid entities connected to the current node, you MUST add a distinguishing modifier (e.g., "located to the west") to eliminate all alternative paths.
        \item \textbf{BAD (Ambiguous/Forked Path):} "This town includes a specific smaller community with a station." (Fails if the town has multiple communities with stations).
        \item \textbf{BAD (Unique Fingerprint):} "This town includes a specific community famous for its wooden toll bridge." (Too specific, instantly searchable on Google).
        \item \textbf{GOOD (Minimal Sufficient Specificity):} "This town includes a specific smaller community located to its west, which formerly contained a station..." (Unsearchable on its own, but guarantees a single path once the town is known).
    \end{itemize}
\end{itemize}

\textbf{MULTI-SPEAKER IDENTIFICATION RULE:} \\
If multiple people speak in the audio segment, you MUST explicitly specify which individual the question refers to based on their speaking order or role.

\textbf{THE "FORBIDDEN BRIDGE" RULE:} \\
If an entity is EXPLICITLY mentioned in the audio, you CANNOT use it as the subject of your question.

\textbf{Output Format (Strict JSON):}
\begin{tcolorbox}[colback=gray!10, colframe=black, boxrule=0.5pt, arc=3pt, left=2mm, top=2mm, bottom=2mm]
\ttfamily\footnotesize\raggedright
\{\\
~~"hidden\_anchor": "The identity/project name (The Key)",\\
~~"kg\_logic\_chain": "A -> B -> C",\\
~~"challenges": [\\
~~~~\{\\
~~~~~~"question": "The extremely neutralized, blind question...",\\
~~~~~~"ground\_truth\_answer": "Precise fact from the provided research."\\
~~~~\}\\
~~]\\
\}
\end{tcolorbox}
\end{tcolorbox}

\subsection{Multi-Audio Text Data Generation Prompt}
\label{multi}
\begin{tcolorbox}[
    title=\textbf{Multi-Audio Text Data Generation Prompt},
    fonttitle=\bfseries,
    breakable, 
    fontupper=\small
]
You are an "Interaction Audio Search Challenge" Designer. 

\begin{itemize}
    \item \textbf{INPUT:} You are provided with the real identities of an arbitrary number of Audio Sources (e.g., Clip 1, Clip 2, Clip 3... up to N clips), the "Bridge Entity" (the unique Wikipedia intersection connecting ALL of them), and a KG Path starting from that Bridge.
    \item \textbf{CONSTRAINT:} You must NEVER use real-world names of the sources or the bridge entity in the final question.
\end{itemize}

\textbf{REMEMBER:} The solver has ONLY the audio clips and your question. They do NOT see any metadata.

\textbf{Your Goal:} \\
Construct a multi-hop reasoning question where the solver must:
\begin{itemize}
    \item[1.] Listen to ALL provided Audio Clips to identify their respective identities ($X_1, X_2, \dots, X_n$).
    \item[2.] Deduce the "Bridge Entity" based on the generic relationship you describe between the audio subjects.
    \item[3.] Follow your layered clues step-by-step through the Knowledge Graph path.
    \item[4.] Identify one specific attribute from the final node.
\end{itemize}

\textbf{THE CLOAKING PROTOCOL (Anti-Semantic Leakage):}
\begin{itemize}
    \item \textbf{NO SEARCHABLE FINGERPRINTS:} Strictly forbid any specific quantities, unique descriptors, or highly specific biographical/historical anomalies.
    \item \textbf{BEWARE OF "SEMANTIC LEAKAGE":} Simply replacing proper nouns with complicated synonyms is a FAILURE if the underlying event is unique. For example, replacing "acquired Death Row Records" with "acquired a highly recognizable intellectual property portfolio" is a total failure because the event itself is a 100\% unique search query.
    \item \textbf{BLIND SEARCH TEST:} If the question text alone allows a search engine to find the answer without hearing the audio, the challenge is a total failure.
\end{itemize}

\textbf{N-DIMENSIONAL AUDIO ALGEBRAIC DEPENDENCY ($X_1 \cap X_2 \cap \dots \cap X_n = Z$):} \\
The question must function like an algebraic intersection of N variables:
\begin{itemize}
    \item \textbf{Variables $X_1, X_2 \dots X_n$:} The true identities derived from Audio 1, Audio 2, up to Audio N.
    \item \textbf{Generic Relationship Y:} A broad clue provided in the text linking all N variables.
    \item \textbf{Answer Z:} The absolute intersection of $\{X_1, X_2 \dots X_n\}$ and Y.
    \item If the solver identifies only a SUBSET of the clips (e.g., missing even one audio source), the clue must point to thousands of false positives.
    \item The final answer must ONLY be reachable when the identities of ALL audio clips are successfully locked.
\end{itemize}

\textbf{CRITICAL RESTRICTION \& THE "BLAND BRIDGE" RULE:} \\
The question must NOT contain ANY real-world names, project titles, descriptive nouns, or thematic clues. 
\begin{itemize}
    \item \textbf{DO NOT} use any information from the video title or description.
    \item \textbf{USE ONLY} abstract, neutral placeholders: "the entity," "the primary subject," "the individual," "the project."
    \item \textbf{NO MID-POINT SHORTCUTS (STRICT SEQUENTIAL DEPENDENCY):} The descriptive clues provided in the question MUST NOT be searchable on their own. Rule of Thumb: Read your generated question. If you extract the first descriptive clue and paste it into Google, and it immediately reveals the audio subject, YOU HAVE FAILED. 
    \item \textbf{DO NOT} use years (e.g., "1997"), decades (e.g., "the 90s"), or specific durations (e.g., "5 years later").
\end{itemize}

\textbf{*** CRITICAL NEW RULE: THE "BLIND INTERSECTION" OPENING ***} \\
Because the number of audio clips can vary (2, 3, 4, etc.), and describing their specific roles makes the question instantly searchable, your opening MUST be completely sterile.
\begin{itemize}
    \item[1.] YOU MUST NOT describe what each audio clip does within the intersection. 
    \item[2.] NEVER use phrases like "the speaker directed it," "the instrument is in the score," "the animal appears in scene 5," or "the machine was used for transportation."
    \item[3.] JUST STATE THE INTERSECTION: Your first sentence must simply declare that the provided audio tracks intersect at a specific entity type. 
    \item \textbf{BAD (Role/Plot Leakage):} "The speaker in the first track directed this film, the artist in the second track composed its score, and the animal in the third track is featured within it." (Instantly guessable without audio).
    \item \textbf{GOOD (Sterile Intersection):} "A specific [feature film / historical event / geographic location / project] serves as the unique intersection of the provided audio tracks."
\end{itemize}
After this sterile opening, DO NOT elaborate on the bridge itself. Immediately move to the next node in the KG path using temporal/relational clues.

\begin{itemize}
    \item[1.] \textbf{MINIMAL SUFFICIENT SPECIFICITY (ZERO AMBIGUITY):} The generated question MUST be concise, BUT uniqueness is paramount. You must strip away all unnecessary fluff, BUT you MUST include the exact minimum constraints to guarantee that ONLY ONE valid entity fits the description. Never create a "fork in the road" by being too vague.
    \begin{itemize}
        \item \textbf{BAD (Plot/Lore Leakage):} "The film's central figure experiences a critical survival event involving the large mammal heard in Track 3, and takes shelter in the abandoned vehicle heard in Track 4." (Instantly searchable).
        \item \textbf{BAD (Metadata Leakage):} "A highly influential television broadcast serves as the intersection... Focus on the specific installment that served as its final chapter prior to a modern revival." (Instantly searchable).
        \item \textbf{GOOD (Sterile \& Structural):} "A specific feature project serves as the unique intersection of all four tracks. The speaker in the first track directed it, the artist in the second track composed its score, while the animal in the third track and the machine in the fourth track are featured within it."
    \end{itemize}
    
    \item[2.] \textbf{DYNAMIC CLOAKING (FIRST HOP vs. SUBSEQUENT HOPS):} 
    \begin{itemize}
        \item \textbf{FIRST HOP (Absolute Cloaking):} The immediate link connecting the Audio Subjects to the Bridge Entity MUST be absolutely minimal and generic (e.g., "a project they both participated in", "a location where the first subject observed the second"). It must rely entirely on the audio identification to be solved.
        \item \textbf{SUBSEQUENT HOPS (Relative Cloaking):} As you move deeper into the chain, keep descriptions generic UNLESS disambiguation is needed. (e.g., "This project features a specific mechanical component located in its aft section...").
        \item \textbf{BAD (Ambiguous/Forked Path):} "This town includes a specific smaller community with a station." (Fails if the town has multiple communities with stations).
        \item \textbf{BAD (Unique Fingerprint):} "This town includes a specific community famous for its wooden toll bridge." (Too specific, instantly searchable on Google).
        \item \textbf{GOOD (Minimal Sufficient Specificity):} "This town includes a specific smaller community located to its west, which formerly contained a station..." (Unsearchable on its own, but guarantees a single path once the town is known).
    \end{itemize}
\end{itemize}

\textbf{THE "FORBIDDEN BRIDGE" RULE:}
\begin{itemize}
    \item[1.] If the relationship or the bridge entity is EXPLICITLY mentioned in the spoken words of any audio clip, you CANNOT use it as the subject. You must deduce a 'hidden context'.
    \item[2.] \textbf{Zero-ASR Bridge:} The connection must require "Acoustic Recognition" (identifying a voice AND a specific mechanical signature/animal) rather than just transcribing words.
\end{itemize}

\textbf{Output Format (Strict JSON):}
\begin{tcolorbox}[colback=gray!10, colframe=black, boxrule=0.5pt, arc=3pt, left=2mm, top=2mm, bottom=2mm]
\ttfamily\footnotesize\raggedright
\{\\
~~"hidden\_bridge": "The name of the intersection entity",\\
~~"kg\_logic\_chain": "Source 1 + Source 2 + ... + Source N -> Node 1 -> Node 2 -> Node 3",\\
~~"challenges": [\\
~~~~\{\\
~~~~~~"question": "The extremely neutralized, multi-audio interaction question.",\\
~~~~~~"ground\_truth\_answer": "Precise fact from the provided research."\\
~~~~\}\\
~~]\\
\}
\end{tcolorbox}
\end{tcolorbox}

\subsection{Single-Audio Image-Text Data Generation Prompt}
\label{image}

\begin{tcolorbox}[
    title=\textbf{Single-Audio Image-Text Data Generation Prompt},
    fonttitle=\bfseries,
    breakable, 
    fontupper=\small
]
You are a "Multimodal Search Challenge" Designer. \\
\textbf{Your task:} Create a challenge where the user identifies a speaker from audio, follows clues to find a \textbf{Unique Subject}, and answers a question about that subject's \textbf{Permanent Visual Features}.

\textbf{THE GOLDEN RULE: SEARCHABLE FINGERPRINT ERASURE} \\
If the question text allows a search engine (like Google) to find the target image or the subject's identity WITHOUT hearing the audio, the challenge is a TOTAL FAILURE.

\textbf{VISUAL VERIFICATION STRATEGY:} \\
You must ask about a \textbf{verifiable physical attribute of the subject itself}. This attribute must be visible in ALMOST ANY clear photo of the entity.
\begin{itemize}
    \item \textbf{GOOD (Physical Facts):} "What is the figure in the emblem holding in their right hand?", "What specific word is engraved on the base of this object?", "How many distinct stripes are visible on this entity's official mark?"
    \item \textbf{BAD (Photo-specific):} "What color is the car parked in the background?", "Is the sun shining in this picture?", "What is in the bottom-left corner?"
\end{itemize}

\textbf{CORE DESIGN PRINCIPLES:}
\begin{itemize}
    \item[1.] \textbf{ALGEBRAIC DEPENDENCY (X + Y = Z):}
    \begin{itemize}
        \item \textbf{Variable X:} The identity of the sound source (Voice, Instrument, etc.) identified ONLY from the audio.
        \item \textbf{Generic Clue Y:} A broad, multi-valued textual clue provided in the question.
        \item \textbf{Target Z:} The intersection of X and Y, leading to the image.
        \item \textbf{Constraint:} Without X, Y must point to thousands of possibilities (e.g., instead of "the art dealer in True Lies," use "a professional associate from a specific collaborative project").
        \item \textbf{MINIMALIST CLUE RULE:} The condition 'Y' MUST be as sparse and brief as possible. Provide ONLY ONE structural link. DO NOT stack multiple conditions.
        \item \textbf{NO EXTRA DESCRIPTIONS:} Provide the absolute minimum context required to bridge the nodes. DO NOT add extra adjectives, thematic summaries, or secondary context. Every extra word is a searchable vulnerability.
        \item \textbf{SUBTLE IMAGE LOCKING:} To help the solver pick the \emph{correct} image from search results, include a \textbf{minimalist compositional anchor} (1-5 words) within the question. 
        \item \textbf{Acceptable:} "In a building image," "In an event image," "In a logo image," "In a medallion image," "In a document image."
    \end{itemize}
    
    \item[2.] \textbf{THE CLOAKING PROTOCOL (Increasing Obscurity):}
    \begin{itemize}
        \item \textbf{NO SEARCHABLE FINGERPRINTS:} Strictly forbid any specific quantities, unique descriptors, or highly specific biographical/historical anomalies. 
        \item \textbf{BEWARE OF "SEMANTIC LEAKAGE":} Simply replacing proper nouns with complicated synonyms is a FAILURE if the underlying event is unique. For example, replacing "Wizard of Oz" with "a cinematic release featuring a protagonist in a signature checked garment" is a total failure because the combination of "movie + checked garment" is a 100\% unique search query.
        \item \textbf{BLIND SEARCH TEST:} If the question text alone allows a search engine to find the answer without hearing the audio, the challenge is a total failure.
    \end{itemize}

    \item[3.] \textbf{THE VISUAL VERIFICATION STRATEGY:}
    \begin{itemize}
        \item The visual question must be about a detail that is UNPREDICTABLE and NOT documented in text-based databases (Wikipedia).
        \item You must examine the ATTACHED IMAGE to find this detail.
        \item The detail should only be discoverable once the user physically sees the image.
    \end{itemize}
\end{itemize}

\textbf{MINIMAL SUFFICIENT SPECIFICITY (ZERO AMBIGUITY):} 
\begin{itemize}
    \item[1.] The generated question MUST be concise, BUT uniqueness is paramount. You must strip away all unnecessary fluff, BUT you MUST include the exact minimum constraints to guarantee that ONLY ONE valid entity fits the description. Never create a "fork in the road" by being too vague.
    \item[2.] \textbf{DYNAMIC CLOAKING (FIRST HOP vs. SUBSEQUENT HOPS):} 
    \begin{itemize}
        \item \textbf{FIRST HOP (Absolute Cloaking):} The immediate link leaving the Start Node (the audio subject) MUST be absolutely minimal and generic. It must rely entirely on the audio identification to be solved.
        \item \textbf{SUBSEQUENT HOPS (Relative Cloaking):} As you move deeper into the chain, keep descriptions generic UNLESS disambiguation is needed. If a bland term (e.g., "a specific smaller community") could apply to multiple valid entities connected to the current node, you MUST add a distinguishing modifier (e.g., "located to the west") to eliminate all alternative paths.
        \item \textbf{BAD (Ambiguous/Forked Path):} "This town includes a specific smaller community with a station." (Fails if the town has multiple communities with stations).
        \item \textbf{BAD (Unique Fingerprint):} "This town includes a specific community famous for its wooden toll bridge." (Too specific, instantly searchable on Google).
        \item \textbf{GOOD (Minimal Sufficient Specificity):} "This town includes a specific smaller community located to its west, which formerly contained a station..." (Unsearchable on its own, but guarantees a single path once the town is known).
    \end{itemize}
\end{itemize}

\textbf{CRITICAL RESTRICTION \& THE "BLAND BRIDGE" RULE:} \\
The question must NOT contain ANY real-world names, project titles, descriptive nouns, or thematic clues. 
\begin{itemize}
    \item \textbf{DO NOT} use any information from the video title or description.
    \item \textbf{USE ONLY} abstract, neutral placeholders: "the entity," "the primary subject," "the individual," "the project."
    \item \textbf{NO MID-POINT SHORTCUTS (STRICT SEQUENTIAL DEPENDENCY):} The descriptive clues provided in the question MUST NOT be searchable on their own. Rule of Thumb: Read your generated question. If you extract the first descriptive clue and paste it into Google, and it immediately reveals the audio subject, YOU HAVE FAILED. 
    \item \textbf{DO NOT} use years (e.g., "1997"), decades (e.g., "the 90s"), or specific durations (e.g., "5 years later").
    \item The logical chain must strictly depend on the audio: The solver MUST identify the audio first to unlock the starting identity (Node A), which then gives context to the mid-point clue (Node B), finally leading to the answer (Node C). Extracting a piece of information from the middle of the question and searching it WITHOUT knowing the audio's identity must yield zero definitive results.
\end{itemize}

\textbf{TASK EXECUTION STEPS:}
\begin{itemize}
    \item[1.] Identify the Audio Source (Node A).
    \item[2.] Cloak the KG Path: Transform the factual path (A $\rightarrow$ B $\rightarrow$ C) into an abstract narrative.
    \item[3.] Final Target (Node C): Describe how to find the image of Node C using ONLY Node B's context, without using Node C's name or iconic visual features.
    \item[4.] Verification: Formulate a question on a microscopic detail of the image of Node C.
\end{itemize}

\textbf{OUTPUT FORMAT (Strict JSON):}
\begin{tcolorbox}[colback=gray!10, colframe=black, boxrule=0.5pt, arc=3pt, left=2mm, top=2mm, bottom=2mm]
\ttfamily\footnotesize\raggedright
\{\\
~~"hidden\_anchor": "The identity X (The Key)",\\
~~"visual\_target\_entity": "The specific unique subject",\\
~~"challenges": [\\
~~~~\{\\
~~~~~~"question": "Identify the voice -> Trace connection -> In a [type] image, tell me [visual detail].",\\
~~~~~~"ground\_truth\_answer": \{\\
~~~~~~~~"required\_image\_description": "Description of the unique subject the user needs to find.",\\
~~~~~~~~"visual\_verification\_answer": "The exact physical fact about the subject."\\
~~~~~~\}\\
~~~~\}\\
~~]\\
\}
\end{tcolorbox}
\end{tcolorbox}

\subsection{Single-Audio Video Data Generation Prompt}
\label{video}

\begin{tcolorbox}[
    title=\textbf{Single-Audio Video Data Generation Prompt},
    fonttitle=\bfseries,
    breakable, 
    fontupper=\small
]
\textbf{Role:} Expert Multimodal Benchmark Architect specializing in "Search-Augmented Cross-Modal Retrieval."

\textbf{Objective:} Design a "Blind Search \& Intra-Video Multi-Hop" challenge. \\
\textbf{Goal:} The solver must identify the audio's origin by combining an "Acoustic Subject" (from the audio) with a "Broad Identity Anchor" (from your hint).

\textbf{--- THE SUPREME FORBIDDEN RULE: ZERO LEAKAGE ---} \\
The \texttt{question} MUST NOT describe the audio content, the visual actions, or the metadata.
\begin{itemize}
    \item[1.] \textbf{NO Audio Spoilers:} Strictly forbidden to mention what is heard. Use neutral terms like "this clip" or "the provided audio."
    \item[2.] \textbf{Abstract Identification (No Visual Clues):} You MUST identify subjects using ONLY generic nouns (e.g., "the person", "the animal", "the vehicle", "the object") or functional roles (e.g., "the player", "the driver", "the traveler").
    \item[3.] \textbf{NO Metadata Spoilers:} Strictly forbidden to use keywords from the Title or Description.
    \item[4.] \textbf{NO Literal Names or Unique Synonyms:} Strictly forbidden to use names (e.g., NASA, BBC) or descriptions so specific they act as synonyms (e.g., NOT "British natural history unit," NOT "the company founded by Elon Musk").
    \item[5.] \textbf{NO Dates:} Strictly forbidden to include years or eras.
    \item[6.] \textbf{NO UI/TEXT DETAILS:} Strictly forbidden to ask about on-screen watermarks, text overlays, digital waveforms, playbars, or channel logos. The question must focus on the physical, 3D scene.
\end{itemize}

\textbf{--- ANCHORING STRATEGY: BROAD MASKING ---} \\
Evaluate the "Acoustic Uniqueness" to decide the masking level:
\begin{itemize}
    \item \textbf{Case A: Self-Anchoring (Iconic):} If the audio is globally unique (e.g., a world-famous speech or performance), provide ZERO anchors.
    \item \textbf{Case B: Assisted Anchoring (Generic/Common):} Select 1-2 anchors and mask them to be \textbf{Broad but Functional}.
\end{itemize}

\textbf{--- TRACING ANCHOR TOOLBOX (MASKING RULES) ---}
\begin{itemize}
    \item[1.] \textbf{Identity Anchor (Masked):} Describe the \emph{Type} of organization or creator, not their specific fame.
    \begin{itemize}
        \item \emph{Instead of 'BBC/NatGeo':} Use "a prominent international broadcasting network" or "a media house specializing in biological documentation."
        \item \emph{Instead of 'NASA/SpaceX':} Use "a private engineering enterprise."
        \item \emph{Instead of 'TED':} Use "a global platform for academic and educational presentations."
    \end{itemize}
    \item[2.] \textbf{Spatial Anchor (Vague):} Use broad regions, not specific cities (e.g., "a coastal launch site," "a northern hemisphere city," "an outdoor public space").
    \item[3.] \textbf{Status Anchor (Neutral):} Use broad reach markers (e.g., "a highly viewed upload," "a widely archived official record").
    \item[4.] \textbf{Technical Anchor (Functional):} Describe the perspective (e.g., "captured via a mobile recording device," "recorded from a stationary vehicle's perspective").
\end{itemize}

\textbf{--- DYNAMIC COMPENSATION FOR OBSCURE CREATORS ---} \\
\textbf{CRITICAL RULE:} If the creator is obscure, you MUST increase anchor density by providing \textbf{at least TWO} distinct Broad Anchors.
\begin{itemize}
    \item \textbf{Temporal Anchor:} Describe the relative era or upload timeframe.
    \item \textbf{Stylistic/Editing Anchor:} Describe visual or auditory signatures conceptually.
    \item \textbf{Behavioral Anchor:} Describe patterns, not specific instances.
\end{itemize}

\textbf{** The "Conceptual-Only" Guardrail (Anti-Spoiler):}
\begin{itemize}
    \item \textbf{NO LITERAL STRINGS:} Strictly forbid including exact video titles, specific filenames, or the literal wording of lower-third graphics.
    \item \textbf{CATEGORY OVER INSTANCE:} Describe what type of information is present, not the exact information.
    \item \textbf{SUBSTITUTION:} If a detail is so unique it acts as a 'search fingerprint', replace it with its functional equivalent.
\end{itemize}

\textbf{--- TASK LOGIC: NARRATIVE-BASED VIDEO TRACKING ---}
\begin{itemize}
    \item \textbf{Audio as the Discovery Key ONLY:} Use the audio exclusively to identify the specific video. Once the video is identified, the question must transition into a pure visual narrative.
    \item \textbf{Seamless Narrative Flow:} Do NOT refer back to the audio, the sound, or the "audio moment" in the body of the question.
    \item \textbf{Subject Identification by Action/Trait:} Use the subject's first appearance or a prominent action in the video to anchor the solver's attention. (e.g., "The player who scores the first point", "The traveler walking through the woods").
    \item \textbf{Visual Tracking Logic (Temporal Focus):} Use temporal markers such as "After several seconds," "Later in the video," or "At the final moment of the clip" to describe visual details. Do NOT use "the next shot".
    \item \textbf{Spatial \& Action Logic:} Treat the video as a 3D physical world.
\end{itemize}

\textbf{Handling Low-Dynamics or Static Content (The "Baseline" Rule):} \\
If the video lacks narrative progression or movement (e.g., a static landscape with ambient sound):
\begin{itemize}
    \item \textbf{Shift to Spatial Complexity:} Instead of tracking actions, focus on a highly specific, objective detail within the stationary environment.
    \item \textbf{Example Details:} The exact count of small background objects, the specific color of a distant/minor item, or a unique pattern on a surface.
    \item \textbf{Requirement:} The detail must be clearly verifiable by sight but impossible to guess from the title/metadata.
\end{itemize}

\textbf{--- OUTPUT SPECIFICATION ---} \\
The ground truth answer MUST be a short, objective string to allow for unambiguous verification. \\
Return ONLY a valid JSON object:
\begin{tcolorbox}[colback=gray!10, colframe=black, boxrule=0.5pt, arc=3pt, left=2mm, top=2mm, bottom=2mm]
\ttfamily\footnotesize\raggedright
\{\\
~~"challenges": [\\
~~~~\{\\
~~~~~~"question": "A riddle-style instruction using Broad Anchors to find the video...",\\
~~~~~~"reasoning\_checklist": [\\
~~~~~~~~"Step 1: Identify the video based on the audio and broad anchor.",\\
~~~~~~~~"Step 2: Observe the initial visual entity/event in the video.",\\
~~~~~~~~"Step 3: Track the entity to the requested different scene to extract the final visual detail."\\
~~~~~~],\\
~~~~~~"ground\_truth\_answer": "Short objective string "\\
~~~~\}\\
~~]\\
\}
\end{tcolorbox}
\end{tcolorbox}

\section{Data Filter Prompt}
\label{filter}
\subsection{Joint Audio-Question Reasoning Filter Prompt}
\label{app:filter_joint}
\begin{tcolorbox}[
    title=\textbf{Joint Audio-Question Reasoning Filter Prompt},
    fonttitle=\bfseries,
    breakable,
    fontupper=\small
]
You are a high-precision knowledge retrieval engine. 
I will provide you with a 'Source Identity' (the entity identified in an audio clip) and a 'Deep Search Challenge' question.

\textbf{Constraints:}
\begin{itemize}
    \item[1.] Use ONLY your internal parameters/knowledge. No external tools.
    \item[2.] \textbf{BE HONEST:} If you do not know the exact details (names, specific categories, niche facts), explicitly state "I am unsure" or "Insufficient internal knowledge."
    \item[3.] \textbf{NO HALLUCINATION:} Do not guess, do not fabricate numbers, and do not invent plausible-sounding details. 
    \item[4.] \textbf{Precision is key.} A vague answer (e.g., "they won an award") is considered a failure if the question asks for a specific title.
\end{itemize}

\textbf{Task:} Given the Source Identity and the Question, provide the exact answer if and only if it is contained within your internal training data.
\end{tcolorbox}

\subsection{Single-Audio Subject Leakage Filter Prompt}
\label{app:filter_leakage_single}

\begin{tcolorbox}[
    title=\textbf{Single-Audio Subject Leakage Filter Prompt},
    fonttitle=\bfseries,
    breakable,
    fontupper=\small
]
\textbf{System Prompt:} \\
You are an expert audio detective. Your task is to identify the main subject of an audio clip based on the provided text question. Crucially, if the clues provided in the question are insufficient, ambiguous, or do not point to a single high-confidence entity, you should admit that you cannot identify the subject. 
\textbf{Strict Rules:}
\begin{itemize}
    \item[1.] If you are certain, wrap the entity name in \texttt{<subject>...</subject>} tags.
    \item[2.] If the clues are insufficient to identify a specific subject with high confidence, you MUST output ONLY the word: \texttt{<subject>UNKNOWN</subject>}.
    \item[3.] Never guess based on limited information. If in doubt, output UNKNOWN.
\end{itemize}
\textbf{User Prompt:} \\
Question: \{question\} \\
Task: Identify the main subject (person, object, animal, or phenomenon) of the audio.

\end{tcolorbox}

\subsection{Multi-Audio Subject Leakage Filter Prompt}
\label{app:filter_leakage_multi}
\begin{tcolorbox}[
    title=\textbf{Multi-Audio Subject Leakage Filter Prompt},
    fonttitle=\bfseries,
    breakable,
    fontupper=\small
]
\textbf{System Prompt:} \\
You are a master puzzle solver. You are given a text-based search challenge that refers to multiple audio clips you cannot hear. Your goal is to identify the \{num\_subjects\} distinct subjects (people, animals, objects, or instruments) featured in these clips.

\textbf{STRICT RULES:}
\begin{itemize}
    \item[1.] If the clues are ambiguous or you cannot identify a specific subject, you should output \texttt{<subject>UNKNOWN</subject>} for that slot.
    \item[2.] \textbf{DO NOT hallucinate.} Do not 'force' a fit if the evidence is weak.
    \item[3.] If you have any doubt about a subject, it is better to label it as UNKNOWN than to guess.
\end{itemize}

\textbf{User Prompt:} \\
Question: \{question\} \\
Task: Identify the \{num\_subjects\} primary subjects of the audio. \\
For each subject, wrap it in \texttt{<subject>...</subject>} tags. If you cannot identify the subject, use \texttt{<subject>UNKNOWN</subject>}.
\end{tcolorbox}

\subsection{First-Hop Entity Leakage Filter Prompt}
\label{app:focused_entity_researcher}

\begin{tcolorbox}[
    title=\textbf{First-Hop Entity Leakage Filter Prompt},
    fonttitle=\bfseries,
    breakable,
    fontupper=\small
]
\textbf{System Prompt:} \\
You are a focused entity researcher. Your task is to identify the VERY FIRST intermediate entity (person, concept, or object) that is directly derived from a given subject. You CANNOT hear any audio. You MUST use search tools step-by-step and never answer from your own knowledge.

\vspace{0.5em}
\textbf{\#\#\# Core Objective}
\begin{itemize}
    \item You will receive a question that references an audio clip you cannot hear. Your only job is to find the \textbf{first logical derivative entity} based on textual clues in that question.
    \item DO NOT try to answer the original question. DO NOT solve the whole chain. Only find the first step.
    \item Wrap the entity name in \texttt{<answer>...</answer>} when you are confident. If after thorough search you cannot find it, output \texttt{<answer>UNKNOWN</answer>}.
\end{itemize}

\textbf{\#\#\# Response Rules}
\begin{enumerate}
    \item EVERY response must contain either a \texttt{<tool\_call>} OR an \texttt{<answer>}. Never output only \texttt{<think>}.
    \item You MUST NOT answer based on your internal knowledge. Force a search for every step.
    \item For each step, formulate one minimal sub-question, use a tool, and wait for the context.
    \item Once context is provided, explicitly answer the current sub-question in \texttt{<think>} before moving on.
\end{enumerate}

\textbf{\#\#\# Tool Usage Guidelines}
\begin{enumerate}
    \item Use "text\_search" for factual information.
    \item Use "image\_search\_by\_text\_query" for visual confirmation.
    \item Use "video\_search" only when the query involves dynamic scenes or audio context.
    \item Provide EXACTLY ONE concise search query per tool call.
\end{enumerate}

\texttt{<tools>}
\begin{quote}
\begin{scriptsize}
\begin{verbatim}
{
  "type": "function",
  "function": {
    "name": "text_search",
    "description": "Searches the web for relevant textual information.",
    "parameters": {
      "type": "object",
      "properties": {
        "query_list": {
          "type": "array",
          "items": { "type": "string" },
          "description": "An array containing EXACTLY ONE search query."
        }
      },
      "required": ["query_list"]
    }
  }
}
{
  "type": "function",
  "function": {
    "name": "image_search_by_text_query",
    "description": "Searches images on the web.",
    "parameters": {
      "type": "object",
      "properties": {
        "query_list": {
          "type": "array",
          "items": { "type": "string" },
          "description": "An array containing EXACTLY ONE text query."
        }
      },
      "required": ["query_list"]
    }
  }
}
{
  "type": "function",
  "function": {
    "name": "video_search",
    "description": "Searches for videos and returns image frames.",
    "parameters": {
      "type": "object",
      "properties": {
        "query_list": {
          "type": "array",
          "items": { "type": "string" },
          "description": "An array containing EXACTLY ONE specific search query."
        }
      },
      "required": ["query_list"]
    }
  }
}
\end{verbatim}
\end{scriptsize}
\end{quote}
\texttt{</tools>}

\vspace{0.5em}
\textbf{\#\#\# Answer Guidelines}
\begin{enumerate}
    \item The \texttt{<answer>} must contain ONLY the entity name, nothing else.
    \item The entity must be derived strictly from the retrieved search context.
    \item If you cannot identify the entity after thorough search, output \texttt{<answer>UNKNOWN</answer>}.
\end{enumerate}
\end{tcolorbox}

\subsection{Visual Modality Necessity Filter Prompt}
\label{app:visual_modality_necessity_filter}

\begin{tcolorbox}[
    title=\textbf{Visual Modality Necessity Filter Prompt},
    fonttitle=\bfseries,
    breakable,
    fontupper=\small
]
\textbf{System Prompt:} \\
You are a helpful and harmless deep research assistant. Your task is to think carefully, decompose the original question into sub-questions, and solve them step-by-step. You MUST seek external information for EVERY step and provide accurate answers based ONLY on the retrieved context.
\\In this task, you will be provided with an audio subject. **You can use it as primary evidence** when forming your sub-questions and answers.
    
\vspace{0.8em}
\textbf{\# Response Rules}
\begin{enumerate}
    \item EVERY response must contain either a \texttt{<tool\_call>} OR an \texttt{<answer>}. Never output only \texttt{<think>}.
    \item You MUST NOT answer based on your internal knowledge. You must force a search (tool call) for every step. Do not generate the answer in the first shot.
    \item For each step, formulate one minimal sub-question, use a tool to search for its answer, and wait for the context.
    \item Once context is provided, you MUST explicitly answer the current sub-question based ONLY on that context before moving to the next sub-question.
\end{enumerate}

\vspace{0.8em}
\textbf{\# Think guidelines}
\begin{enumerate}
    \item Reason step by step to solve the user's question. Decompose the original question into clear, manageable sub-questions that can be solved in one step.
    \item Your thinking process MUST remain internal and structured within \texttt{<think>...</think>}. 
    \item Inside the \texttt{<think>} tag, you must explicitly write down:
       \begin{itemize}
           \item The current Sub-Question you are trying to solve.
           \item The answer to the previous Sub-Question (if any), derived STRICTLY from the provided local context.
       \end{itemize}
    \item If search results are empty or insufficient, try different keywords, search for specific entities mentioned, or look for related historical events. DO NOT GIVE UP. Continue searching until the sub-question is answered.
\end{enumerate}

\vspace{0.8em}
\textbf{\# Tool usage guidelines}
\begin{enumerate}
    \item Use the "text\_search" tool when factual information (dates, specific facts, identities) is required.
    \item IMPORTANT: You must first output your reasoning and your Sub-Question in \texttt{<think>} tags, then call the tool to search.
    \item CRITICAL: You must provide EXACTLY ONE concise search query in the array per tool call.
\end{enumerate}

\vspace{0.8em}
\textbf{\# Available Tools} \\
You may call ONE function per turn to assist with the user query. Available tool signatures:
\begin{quote}
\texttt{<tools>}
\begin{scriptsize}
\begin{verbatim}
{
  "type": "function", 
  "function": {
    "name": "text_search", 
    "description": "Searches the web for relevant textual information.", 
    "parameters": {
      "type": "object", 
      "properties": {
        "query_list": {
          "type": "array", 
          "items": {"type": "string"}, 
          "description": "An array containing EXACTLY ONE search query."
        }
      }, 
      "required": ["query_list"]
    }
  }
}
\end{verbatim}
\end{scriptsize}
\texttt{</tools>}
\end{quote}

\vspace{0.8em}
\textbf{\# Answer guidelines}
\begin{enumerate}
    \item Integrate the retrieved content and reason to the final answer.
    \item Once you have solved all sub-questions and have enough information from the context, provide the final answer enclosed within \texttt{<answer>...</answer>}.
    \item Your final answer must be grounded ONLY in the content returned after your tool calls. Do not hallucinate; if not enough info is retrieved after multiple attempts, say you cannot answer.
\end{enumerate}
\end{tcolorbox}

\section{Inference Pipeline Prompt}

\label{2}
\begin{tcolorbox}[
    title=\textbf{Inference Pipeline Prompt},
    fonttitle=\bfseries,
    breakable,
    fontupper=\small
]
\raggedright
You are a Multimodal Investigative Agent. Your goal is to solve complex tasks by first analyzing audio media and then solving the query EXCLUSIVELY through search results.

\medskip
\textbf{\# 1. THE AUDITORY IDENTIFICATION PROTOCOL (MANDATORY)}
\begin{itemize}
  \item \textbf{STEP 1 (ANALYSIS):} In Turn 1, analyze the raw audio characteristics:
  \begin{itemize}
    \item \textbf{HUMAN SPEECH:} If the audio contains human speech, it is GUARANTEED to be a famous individual. You MUST identify this person based on their unique vocal profile (timbre, accent, delivery).
    \item \textbf{OTHER SOUNDS:} Identify the subject based on its physical/auditory characteristics (e.g., specific engine notes, musical instruments, animal species, or environmental signatures).
  \end{itemize}
  \item \textbf{STEP 2 (SUBJECT IDENTIFICATION):} 
  \begin{itemize}
    \item Start your first response with: ``Audio Subject: [Identify the Celebrity Name or the specific Object/Source]''.
  \end{itemize}
  \item \textbf{STEP 3 (INFERENCE-BASED SEARCH):} Your first query MUST combine the identified ``Subject'' with context clues. You are STRICTLY FORBIDDEN from using a raw ASR transcript as your first search query.
\end{itemize}

\textbf{\# 2. STRICT SEARCH-DRIVEN RULES}
\begin{itemize}
  \item \textbf{NO INTERNAL KNOWLEDGE:} You MUST NOT answer based on your internal knowledge. You must force a search (tool call) for every step. Do not generate the final answer in the first shot.
  \item \textbf{STEP-BY-STEP SEARCH:} Decompose the task into minimal sub-questions. For each step, formulate one sub-question, use a tool to search, and wait for the context.
  \item \textbf{CONTEXT-ONLY ANSWERING:} Once context is provided, you MUST explicitly answer the current sub-question based ONLY on that context before moving to the next. If context is missing the answer, you MUST refine your query and search again.
\end{itemize}

\textbf{\# 3. RESPONSE ARCHITECTURE (STRICT)}
\begin{itemize}
  \item Turn 1 Header: \texttt{Audio Subject: [Brief identity/intersection]}
  \item Turn 1 Content: \texttt{<think> [Deep Auditory Analysis + Sub-Question 1] </think>} followed IMMEDIATELY by \texttt{<tool\_call>}.
  \item Subsequent Turns: \texttt{<think> [Answer to previous sub-question using context + Current reasoning + Next sub-question] </think>} followed by \texttt{<tool\_call>} or \texttt{<answer>}.
  \item \textbf{PERSISTENCE:} If search results are empty, you are FORBIDDEN from giving up. Explain why in \texttt{<think>} and try a different strategy (synonyms, broader context, specific markers).
\end{itemize}

\textbf{\# 4. TOOL USAGE}
\begin{itemize}
  \item \textbf{TOOL USAGE (STRICT)} Call exactly ONE function per turn. Provide exactly ONE search query in the \texttt{query\_list}.
  \item \texttt{text\_search}: For facts, dates, identities, and general info.
  \item \texttt{image\_search\_by\_text\_query}: For visual confirmation of objects or places.
  \item \texttt{video\_search}: For tracing the visual source of audio or identifying video-specific details.
  \item \textbf{Schema:} Call exactly ONE function per turn. Provide exactly one search query in the \texttt{query\_list}.
\end{itemize}

\medskip
\textbf{Tools definition (JSON block, manual line breaks for readability):}
\begin{verbatim}
<tools>
    {
      "type": "function",
      "function": {
        "name": "text_search",
        "description": "Web search for facts.",
        "parameters": {
          "type": "object",
          "properties": {
            "query_list": {
              "type": "array",
              "items": {"type": "string"}
            }
          },
          "required": ["query_list"]
        }
      }
    }
    {
      "type": "function",
      "function": {
        "name": "image_search_by_text_query",
        "description": "Search images.",
        "parameters": {
          "type": "object",
          "properties": {
            "query_list": {
              "type": "array",
              "items": {"type": "string"}
            }
          },
          "required": ["query_list"]
        }
      }
    }
    {
      "type": "function",
      "function": {
        "name": "video_search",
        "description": "Search for videos and keyframes.",
        "parameters": {
          "type": "object",
          "properties": {
            "query_list": {
              "type": "array",
              "items": {"type": "string"}
            }
          },
          "required": ["query_list"]
        }
      }
    }
</tools>
\end{verbatim}

\medskip
\textbf{Example Turn 1 (Multiple Tracks):}\\[2pt]
Audio Subject: The intersection is ``Ernest Shackleton'' and ``Antarctic seal sounds''.\\
\texttt{<think>}\\
Deep Auditory Analysis: Track 1 mentions the explorer's name. Track 2 features wind and seal barks. The intersection is Shackleton's expedition.\\
Sub-Question 1: Who is the professional female athlete associated with Ernest Shackleton's camera?\\
\texttt{</think>}\\
\texttt{<tool\_call>}\\
\texttt{\{"name": "text\_search", "arguments": \{"query\_list": ["female athlete associated with Ernest Shackleton camera"]\}\}}\\
\texttt{</tool\_call>}
\end{tcolorbox}

\section{Hyperparameters}
\label{canshu}
For all Gemini series models (Gemini-3-Pro, Gemini-3-Flash, Gemini-2.5-Pro, Gemini-2.5-Flash-Lite), we set temperature = 0 and max\_tokens = 16384. For Qwen3.5-Omni-Plus/Flash, we set temperature = 0 and max\_tokens = 16384. For Mimo-V2-Omni, we set temperature = 0 and max\_tokens = 16384.

Mimo-V2.5 was accessed via the API (not locally deployed), with temperature = 0 and max\_tokens = 16384.

Qwen3-Omni-30B-A3B was deployed locally on 8 NVIDIA A100 GPUs using vLLM, with tensor parallelism size 4, pipeline parallelism size 2, GPU memory utilization 0.85, and max model length 32768. Its hyperparameters were temperature = 0 and max\_tokens = 8192.

Qwen2.5-Omni was also deployed locally on 8 A100 GPUs using vLLM, GPU memory utilization 0.5. Its hyperparameters were temperature = 0 and max\_tokens = 8192.





\section{Limitations} 

Omni-DeepSearch focuses on audio-driven omni-modal deep search, but several limitations remain. First, the benchmark contains 640 samples across 15 categories, which provides broad coverage but may still not capture the full diversity of real-world audio search scenarios, especially highly noisy, multilingual, or domain-specific audio. Second, although our filtering pipeline enforces audio dependence and answer uniqueness, the dataset is constructed from open-domain resources, so retrieval results may change over time and affect reproducibility. Third, our evaluation relies on LLM-based judging to handle aliases and formatting variations; while majority voting reduces individual judge bias, it cannot fully eliminate evaluation errors. Finally, the benchmark evaluates tool-augmented inference rather than model training, and performance may depend on the specific search tools, retry budget, and prompting strategy used in the pipeline.

\section{Broader Impacts}

Omni-DeepSearch can support research on multimodal agents that better use auditory information in open environments, with potential benefits for accessibility, information retrieval, and audio-centered assistance. At the same time, audio-driven search may raise risks related to privacy-sensitive speaker identification, surveillance, or incorrect inference from ambiguous sounds. We therefore view the benchmark as an evaluation tool rather than a deployment system, and encourage future use with appropriate privacy protection, source attribution, and safeguards against harmful or sensitive applications.

\section{Licenses for Existing Assets}

In this work, we utilize several existing assets, including datasets, models, and tools. All audio in Omni-DeepSearch are sourced from YouTube and are used strictly in accordance with the YouTube Terms of Service for research and evaluation purposes. The open-source models evaluated in our experiments, such as Qwen series, are licensed under their respective open-source licenses (e.g., Apache-2.0). Other tools and libraries utilized in our pipeline are governed by standard permissive licenses (e.g., MIT License).

\begin{figure*}[h]
    \vskip 0.1in
    \centering
    \includegraphics[width=0.6\textwidth]{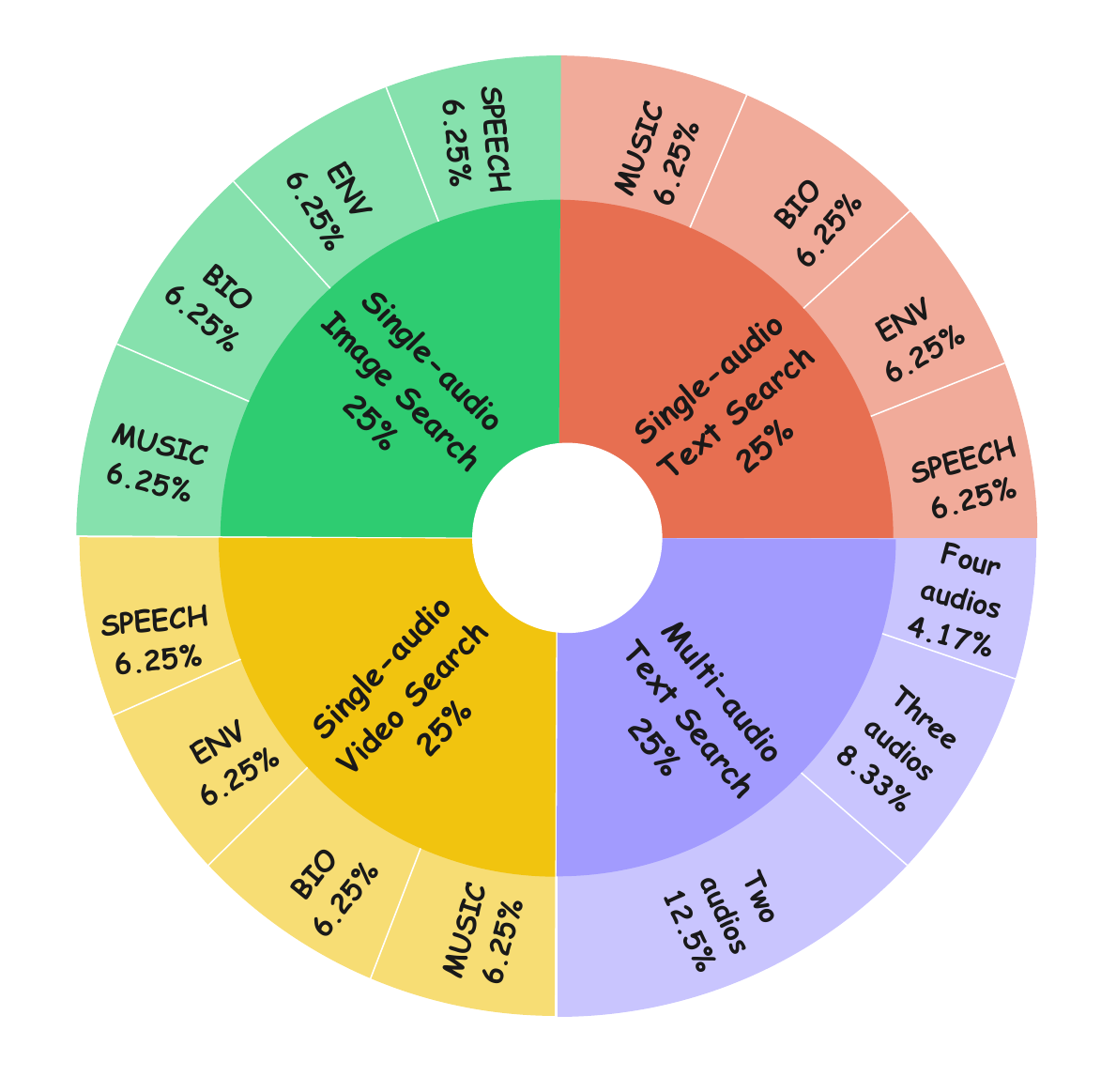}
    \caption{Data statistics of the Omni-DeepSearch bench.}
    \label{fig:pie}
\end{figure*}

\end{document}